\RequirePackage{lineno}

\documentclass[aps,prl,twocolumn,superscriptaddress,showpacs,preprintnumbers,amsmath,amssymb]{revtex4-1}

\usepackage{graphicx} 
\usepackage{dcolumn}  
\usepackage{color} %
\RequirePackage{xspace}
\usepackage{relsize}
\usepackage{colortbl}
\usepackage{amsmath} 
\usepackage{ifthen} 
\usepackage{subfigure}

\usepackage[normalem]{ulem}

\usepackage{ifthen} 
\graphicspath{{ps}}
\newboolean{pdflatex}
\setboolean{pdflatex}{true} 
\usepackage{rotating} 

\newboolean{articletitles}
\setboolean{articletitles}{true} 

\newboolean{uprightparticles}
\setboolean{uprightparticles}{false} 

\usepackage{amssymb}
\usepackage{amsfonts}
\usepackage{upgreek} 

\usepackage{hyperref}    
\usepackage[all]{hypcap} 

\begin{document}



\def\lhcb {LHCb\xspace}
\def\ux85 {UX85\xspace}
\def\cern {CERN\xspace}
\def\lhc {LHC\xspace}
\def\atlas {ATLAS\xspace}
\def\cms {CMS\xspace}
\def\babar  {BaBar\xspace}
\def\belle  {Belle\xspace}
\def\aleph  {ALEPH\xspace}
\def\delphi {DELPHI\xspace}
\def\opal   {OPAL\xspace}
\def\lthree {L3\xspace}
\def\lep    {LEP\xspace}
\def\cdf    {CDF\xspace}
\def\dzero  {D\O\xspace}
\def\sld    {SLD\xspace}
\def\cleo   {CLEO\xspace}
\def\uaone  {UA1\xspace}
\def\uatwo  {UA2\xspace}
\def\tevatron {TEVATRON\xspace}


\def\pu     {PU\xspace}
\def\velo   {VELO\xspace}
\def\rich   {RICH\xspace}
\def\richone {RICH1\xspace}
\def\richtwo {RICH2\xspace}
\def\ttracker {TT\xspace}
\def\intr   {IT\xspace}
\def\st     {ST\xspace}
\def\ot     {OT\xspace}
\def\Tone   {T1\xspace}
\def\Ttwo   {T2\xspace}
\def\Tthree {T3\xspace}
\def\Mone   {M1\xspace}
\def\Mtwo   {M2\xspace}
\def\Mthree {M3\xspace}
\def\Mfour  {M4\xspace}
\def\Mfive  {M5\xspace}
\def\ecal   {ECAL\xspace}
\def\spd    {SPD\xspace}
\def\presh  {PS\xspace}
\def\hcal   {HCAL\xspace}
\def\bcm    {BCM\xspace}

\def\ode    {ODE\xspace}
\def\daq    {DAQ\xspace}
\def\tfc    {TFC\xspace}
\def\ecs    {ECS\xspace}
\def\lone   {L0\xspace}
\def\hlt    {HLT\xspace}
\def\hltone {HLT1\xspace}
\def\hlttwo {HLT2\xspace}


\ifthenelse{\boolean{uprightparticles}}%
{\def\Palpha      {\ensuremath{\upalpha}\xspace}
 \def\Pbeta       {\ensuremath{\upbeta}\xspace}
 \def\Pgamma      {\ensuremath{\upgamma}\xspace}                 
 \def\Pdelta      {\ensuremath{\updelta}\xspace}                 
 \def\Pepsilon    {\ensuremath{\upepsilon}\xspace}                 
 \def\Pvarepsilon {\ensuremath{\upvarepsilon}\xspace}                 
 \def\Pzeta       {\ensuremath{\upzeta}\xspace}                 
 \def\Peta        {\ensuremath{\upeta}\xspace}                 
 \def\Ptheta      {\ensuremath{\uptheta}\xspace}                 
 \def\Pvartheta   {\ensuremath{\upvartheta}\xspace}                 
 \def\Piota       {\ensuremath{\upiota}\xspace}                 
 \def\Pkappa      {\ensuremath{\upkappa}\xspace}                 
 \def\Plambda     {\ensuremath{\uplambda}\xspace}                 
 \def\Pmu         {\ensuremath{\upmu}\xspace}                 
 \def\Pnu         {\ensuremath{\upnu}\xspace}                 
 \def\Pxi         {\ensuremath{\upxi}\xspace}                 
 \def\Ppi         {\ensuremath{\uppi}\xspace}                 
 \def\Pvarpi      {\ensuremath{\upvarpi}\xspace}                 
 \def\Prho        {\ensuremath{\uprho}\xspace}                 
 \def\Pvarrho     {\ensuremath{\upvarrho}\xspace}                 
 \def\Ptau        {\ensuremath{\uptau}\xspace}                 
 \def\Pupsilon    {\ensuremath{\upupsilon}\xspace}                 
 \def\Pphi        {\ensuremath{\upphi}\xspace}                 
 \def\Pvarphi     {\ensuremath{\upvarphi}\xspace}                 
 \def\Pchi        {\ensuremath{\upchi}\xspace}                 
 \def\Ppsi        {\ensuremath{\uppsi}\xspace}                 
 \def\Pomega      {\ensuremath{\upomega}\xspace}                 

 \def\PDelta      {\ensuremath{\Delta}\xspace}                 
 \def\PXi      {\ensuremath{\Xi}\xspace}                 
 \def\PLambda      {\ensuremath{\Lambda}\xspace}                 
 \def\PSigma      {\ensuremath{\Sigma}\xspace}                 
 \def\POmega      {\ensuremath{\Omega}\xspace}                 
 \def\PUpsilon      {\ensuremath{\Upsilon}\xspace}                 
 

 \def\PA      {\ensuremath{\mathrm{A}}\xspace}                 
 \def\PB      {\ensuremath{\mathrm{B}}\xspace}                 
 \def\PC      {\ensuremath{\mathrm{C}}\xspace}                 
 \def\PD      {\ensuremath{\mathrm{D}}\xspace}                 
 \def\PE      {\ensuremath{\mathrm{E}}\xspace}                 
 \def\PF      {\ensuremath{\mathrm{F}}\xspace}                 
 \def\PG      {\ensuremath{\mathrm{G}}\xspace}                 
 \def\PH      {\ensuremath{\mathrm{H}}\xspace}                 
 \def\PI      {\ensuremath{\mathrm{I}}\xspace}                 
 \def\PJ      {\ensuremath{\mathrm{J}}\xspace}                 
 \def\PK      {\ensuremath{\mathrm{K}}\xspace}                 
 \def\PL      {\ensuremath{\mathrm{L}}\xspace}                 
 \def\PM      {\ensuremath{\mathrm{M}}\xspace}                 
 \def\PN      {\ensuremath{\mathrm{N}}\xspace}                 
 \def\PO      {\ensuremath{\mathrm{O}}\xspace}                 
 \def\PP      {\ensuremath{\mathrm{P}}\xspace}                 
 \def\PQ      {\ensuremath{\mathrm{Q}}\xspace}                 
 \def\PR      {\ensuremath{\mathrm{R}}\xspace}                 
 \def\PS      {\ensuremath{\mathrm{S}}\xspace}                 
 \def\PT      {\ensuremath{\mathrm{T}}\xspace}                 
 \def\PU      {\ensuremath{\mathrm{U}}\xspace}                 
 \def\PV      {\ensuremath{\mathrm{V}}\xspace}                 
 \def\PW      {\ensuremath{\mathrm{W}}\xspace}                 
 \def\PX      {\ensuremath{\mathrm{X}}\xspace}                 
 \def\PY      {\ensuremath{\mathrm{Y}}\xspace}                 
 \def\PZ      {\ensuremath{\mathrm{Z}}\xspace}                 
 \def\Pa      {\ensuremath{\mathrm{a}}\xspace}                 
 \def\Pb      {\ensuremath{\mathrm{b}}\xspace}                 
 \def\Pc      {\ensuremath{\mathrm{c}}\xspace}                 
 \def\Pd      {\ensuremath{\mathrm{d}}\xspace}                 
 \def\Pe      {\ensuremath{\mathrm{e}}\xspace}                 
 \def\Pf      {\ensuremath{\mathrm{f}}\xspace}                 
 \def\Pg      {\ensuremath{\mathrm{g}}\xspace}                 
 \def\Ph      {\ensuremath{\mathrm{h}}\xspace}                 
 \def\Pi      {\ensuremath{\mathrm{i}}\xspace}                 
 \def\Pj      {\ensuremath{\mathrm{j}}\xspace}                 
 \def\Pk      {\ensuremath{\mathrm{k}}\xspace}                 
 \def\Pl      {\ensuremath{\mathrm{l}}\xspace}                 
 \def\Pm      {\ensuremath{\mathrm{m}}\xspace}                 
 \def\Pn      {\ensuremath{\mathrm{n}}\xspace}                 
 \def\Po      {\ensuremath{\mathrm{o}}\xspace}                 
 \def\Pp      {\ensuremath{\mathrm{p}}\xspace}                 
 \def\Pq      {\ensuremath{\mathrm{q}}\xspace}                 
 \def\Pr      {\ensuremath{\mathrm{r}}\xspace}                 
 \def\Ps      {\ensuremath{\mathrm{s}}\xspace}                 
 \def\Pt      {\ensuremath{\mathrm{t}}\xspace}                 
 \def\Pu      {\ensuremath{\mathrm{u}}\xspace}                 
 \def\Pv      {\ensuremath{\mathrm{v}}\xspace}                 
 \def\Pw      {\ensuremath{\mathrm{w}}\xspace}                 
 \def\Px      {\ensuremath{\mathrm{x}}\xspace}                 
 \def\Py      {\ensuremath{\mathrm{y}}\xspace}                 
 \def\Pz      {\ensuremath{\mathrm{z}}\xspace}                 
}
{\def\Palpha      {\ensuremath{\alpha}\xspace}
 \def\Pbeta       {\ensuremath{\beta}\xspace}
 \def\Pgamma      {\ensuremath{\gamma}\xspace}                 
 \def\Pdelta      {\ensuremath{\delta}\xspace}                 
 \def\Pepsilon    {\ensuremath{\epsilon}\xspace}                 
 \def\Pvarepsilon {\ensuremath{\varepsilon}\xspace}                 
 \def\Pzeta       {\ensuremath{\zeta}\xspace}                 
 \def\Peta        {\ensuremath{\eta}\xspace}                 
 \def\Ptheta      {\ensuremath{\theta}\xspace}                 
 \def\Pvartheta   {\ensuremath{\vartheta}\xspace}                 
 \def\Piota       {\ensuremath{\iota}\xspace}                 
 \def\Pkappa      {\ensuremath{\kappa}\xspace}                 
 \def\Plambda     {\ensuremath{\lambda}\xspace}                 
 \def\Pmu         {\ensuremath{\mu}\xspace}                 
 \def\Pnu         {\ensuremath{\nu}\xspace}                 
 \def\Pxi         {\ensuremath{\xi}\xspace}                 
 \def\Ppi         {\ensuremath{\pi}\xspace}                 
 \def\Pvarpi      {\ensuremath{\varpi}\xspace}                 
 \def\Prho        {\ensuremath{\rho}\xspace}                 
 \def\Pvarrho     {\ensuremath{\varrho}\xspace}                 
 \def\Ptau        {\ensuremath{\tau}\xspace}                 
 \def\Pupsilon    {\ensuremath{\upsilon}\xspace}                 
 \def\Pphi        {\ensuremath{\phi}\xspace}                 
 \def\Pvarphi     {\ensuremath{\varphi}\xspace}                 
 \def\Pchi        {\ensuremath{\chi}\xspace}                 
 \def\Ppsi        {\ensuremath{\psi}\xspace}                 
 \def\Pomega      {\ensuremath{\omega}\xspace}                 
 \mathchardef\PDelta="7101
 \mathchardef\PXi="7104
 \mathchardef\PLambda="7103
 \mathchardef\PSigma="7106
 \mathchardef\POmega="710A
 \mathchardef\PUpsilon="7107
 \def\PA      {\ensuremath{A}\xspace}                 
 \def\PB      {\ensuremath{B}\xspace}                 
 \def\PC      {\ensuremath{C}\xspace}                 
 \def\PD      {\ensuremath{D}\xspace}                 
 \def\PE      {\ensuremath{E}\xspace}                 
 \def\PF      {\ensuremath{F}\xspace}                 
 \def\PG      {\ensuremath{G}\xspace}                 
 \def\PH      {\ensuremath{H}\xspace}                 
 \def\PI      {\ensuremath{I}\xspace}                 
 \def\PJ      {\ensuremath{J}\xspace}                 
 \def\PK      {\ensuremath{K}\xspace}                 
 \def\PL      {\ensuremath{L}\xspace}                 
 \def\PM      {\ensuremath{M}\xspace}                 
 \def\PN      {\ensuremath{N}\xspace}                 
 \def\PO      {\ensuremath{O}\xspace}                 
 \def\PP      {\ensuremath{P}\xspace}                 
 \def\PQ      {\ensuremath{Q}\xspace}                 
 \def\PR      {\ensuremath{R}\xspace}                 
 \def\PS      {\ensuremath{S}\xspace}                 
 \def\PT      {\ensuremath{T}\xspace}                 
 \def\PU      {\ensuremath{U}\xspace}                 
 \def\PV      {\ensuremath{V}\xspace}                 
 \def\PW      {\ensuremath{W}\xspace}                 
 \def\PX      {\ensuremath{X}\xspace}                 
 \def\PY      {\ensuremath{Y}\xspace}                 
 \def\PZ      {\ensuremath{Z}\xspace}                 
 \def\Pa      {\ensuremath{a}\xspace}                 
 \def\Pb      {\ensuremath{b}\xspace}                 
 \def\Pc      {\ensuremath{c}\xspace}                 
 \def\Pd      {\ensuremath{d}\xspace}                 
 \def\Pe      {\ensuremath{e}\xspace}                 
 \def\Pf      {\ensuremath{f}\xspace}                 
 \def\Pg      {\ensuremath{g}\xspace}                 
 \def\Ph      {\ensuremath{h}\xspace}                 
 \def\Pi      {\ensuremath{i}\xspace}                 
 \def\Pj      {\ensuremath{j}\xspace}                 
 \def\Pk      {\ensuremath{k}\xspace}                 
 \def\Pl      {\ensuremath{l}\xspace}                 
 \def\Pm      {\ensuremath{m}\xspace}                 
 \def\Pn      {\ensuremath{n}\xspace}                 
 \def\Po      {\ensuremath{o}\xspace}                 
 \def\Pp      {\ensuremath{p}\xspace}                 
 \def\Pq      {\ensuremath{q}\xspace}                 
 \def\Pr      {\ensuremath{r}\xspace}                 
 \def\Ps      {\ensuremath{s}\xspace}                 
 \def\Pt      {\ensuremath{t}\xspace}                 
 \def\Pu      {\ensuremath{u}\xspace}                 
 \def\Pv      {\ensuremath{v}\xspace}                 
 \def\Pw      {\ensuremath{w}\xspace}                 
 \def\Px      {\ensuremath{x}\xspace}                 
 \def\Py      {\ensuremath{y}\xspace}                 
 \def\Pz      {\ensuremath{z}\xspace}                 
}



\let\emi\en
\def\electron   {\ensuremath{\Pe}\xspace}
\def\en         {\ensuremath{\Pe^-}\xspace}   
\def\ep         {\ensuremath{\Pe^+}\xspace}
\def\epm        {\ensuremath{\Pe^\pm}\xspace} 
\def\epem       {\ensuremath{\Pe^+\Pe^-}\xspace}
\def\ee         {\ensuremath{\Pe^-\Pe^-}\xspace}

\def\mmu        {\ensuremath{\Pmu}\xspace}
\def\mup        {\ensuremath{\Pmu^+}\xspace}
\def\mun        {\ensuremath{\Pmu^-}\xspace} 
\def\mumu       {\ensuremath{\Pmu^+\Pmu^-}\xspace}
\def\mtau       {\ensuremath{\Ptau}\xspace}

\def\taup       {\ensuremath{\Ptau^+}\xspace}
\def\taum       {\ensuremath{\Ptau^-}\xspace}
\def\tautau     {\ensuremath{\Ptau^+\Ptau^-}\xspace}

\def\ellm       {\ensuremath{\ell^-}\xspace}
\def\ellp       {\ensuremath{\ell^+}\xspace}
\def\ellell     {\ensuremath{\ell^+ \ell^-}\xspace}

\def\neu        {\ensuremath{\Pnu}\xspace}
\def\neub       {\ensuremath{\overline{\Pnu}}\xspace}
\def\nuenueb    {\ensuremath{\neu\neub}\xspace}
\def\neue       {\ensuremath{\neu_e}\xspace}
\def\neueb      {\ensuremath{\neub_e}\xspace}
\def\neueneueb  {\ensuremath{\neue\neueb}\xspace}
\def\neum       {\ensuremath{\neu_\mu}\xspace}
\def\neumb      {\ensuremath{\neub_\mu}\xspace}
\def\neumneumb  {\ensuremath{\neum\neumb}\xspace}
\def\neut       {\ensuremath{\neu_\tau}\xspace}
\def\neutb      {\ensuremath{\neub_\tau}\xspace}
\def\neutneutb  {\ensuremath{\neut\neutb}\xspace}
\def\neul       {\ensuremath{\neu_\ell}\xspace}
\def\neulb      {\ensuremath{\neub_\ell}\xspace}
\def\neulneulb  {\ensuremath{\neul\neulb}\xspace}


\def\g      {\ensuremath{\Pgamma}\xspace}
\def\H      {\ensuremath{\PH^0}\xspace}
\def\Hp     {\ensuremath{\PH^+}\xspace}
\def\Hm     {\ensuremath{\PH^-}\xspace}
\def\Hpm    {\ensuremath{\PH^\pm}\xspace}
\def\W      {\ensuremath{\PW}\xspace}
\def\Wp     {\ensuremath{\PW^+}\xspace}
\def\Wm     {\ensuremath{\PW^-}\xspace}
\def\Wpm    {\ensuremath{\PW^\pm}\xspace}
\def\Z      {\ensuremath{\PZ^0}\xspace}


\def\quark     {\ensuremath{\Pq}\xspace}
\def\quarkbar  {\ensuremath{\overline \quark}\xspace}
\def\qqbar     {\ensuremath{\quark\quarkbar}\xspace}
\def\uquark    {\ensuremath{\Pu}\xspace}
\def\uquarkbar {\ensuremath{\overline \uquark}\xspace}
\def\uubar     {\ensuremath{\uquark\uquarkbar}\xspace}
\def\dquark    {\ensuremath{\Pd}\xspace}
\def\dquarkbar {\ensuremath{\overline \dquark}\xspace}
\def\ddbar     {\ensuremath{\dquark\dquarkbar}\xspace}
\def\squark    {\ensuremath{\Ps}\xspace}
\def\squarkbar {\ensuremath{\overline \squark}\xspace}
\def\ssbar     {\ensuremath{\squark\squarkbar}\xspace}
\def\cquark    {\ensuremath{\Pc}\xspace}
\def\cquarkbar {\ensuremath{\overline \cquark}\xspace}
\def\ccbar     {\ensuremath{\cquark\cquarkbar}\xspace}
\def\bquark    {\ensuremath{\Pb}\xspace}
\def\bquarkbar {\ensuremath{\overline \bquark}\xspace}
\def\bbbar     {\ensuremath{\bquark\bquarkbar}\xspace}
\def\tquark    {\ensuremath{\Pt}\xspace}
\def\tquarkbar {\ensuremath{\overline \tquark}\xspace}
\def\ttbar     {\ensuremath{\tquark\tquarkbar}\xspace}


\def\pion  {\ensuremath{\Ppi}\xspace}
\def\piz   {\ensuremath{\pion^0}\xspace}
\def\pizs  {\ensuremath{\pion^0\mbox\,\rm{s}}\xspace}
\def\ppz   {\ensuremath{\pion^0\pion^0}\xspace}
\def\pip   {\ensuremath{\pion^+}\xspace}
\def\pim   {\ensuremath{\pion^-}\xspace}
\def\pipi  {\ensuremath{\pion^+\pion^-}\xspace}
\def\pipm  {\ensuremath{\pion^\pm}\xspace}
\def\pimp  {\ensuremath{\pion^\mp}\xspace}

\def\kaon  {\ensuremath{\PK}\xspace}
  \def\Kbar  {\kern 0.2em\overline{\kern -0.2em \PK}{}\xspace}
\def\Kb    {\ensuremath{\Kbar}\xspace}
\def\Kz    {\ensuremath{\kaon^0}\xspace}
\def\Kzb   {\ensuremath{\Kbar^0}\xspace}
\def\KzKzb {\ensuremath{\Kz \kern -0.16em \Kzb}\xspace}
\def\Kp    {\ensuremath{\kaon^+}\xspace}
\def\Km    {\ensuremath{\kaon^-}\xspace}
\def\Kpm   {\ensuremath{\kaon^\pm}\xspace}
\def\Kmp   {\ensuremath{\kaon^\mp}\xspace}
\def\KpKm  {\ensuremath{\Kp \kern -0.16em \Km}\xspace}
\def\KS    {\ensuremath{\kaon^0_{\rm\scriptscriptstyle S}}\xspace} 
\def\KSb    {\ensuremath{\Kbar^0_{\rm\scriptscriptstyle S}}\xspace} 
\def\KL    {\ensuremath{\kaon^0_{\rm\scriptscriptstyle L}}\xspace} 
\def\Kstarz  {\ensuremath{\kaon^{*0}}\xspace}
\def\Kstarzb {\ensuremath{\Kbar^{*0}}\xspace}
\def\Kstar   {\ensuremath{\kaon^*}\xspace}
\def\Kstarb  {\ensuremath{\Kbar^*}\xspace}
\def\Kstarp  {\ensuremath{\kaon^{*+}}\xspace}
\def\Kstarm  {\ensuremath{\kaon^{*-}}\xspace}
\def\Kstarpm {\ensuremath{\kaon^{*\pm}}\xspace}
\def\Kstarmp {\ensuremath{\kaon^{*\mp}}\xspace}

\newcommand{\etapr}{\ensuremath{\Peta^{\prime}}\xspace}


  \def\Dbar    {\kern 0.2em\overline{\kern -0.2em \PD}{}\xspace}
\def\D       {\ensuremath{\PD}\xspace}
\def\Db      {\ensuremath{\Dbar}\xspace}
\def\Dz      {\ensuremath{\D^0}\xspace}
\def\Dzb     {\ensuremath{\Dbar^0}\xspace}
\def\DzDzb   {\ensuremath{\Dz {\kern -0.16em \Dzb}}\xspace}
\def\Dp      {\ensuremath{\D^+}\xspace}
\def\Dm      {\ensuremath{\D^-}\xspace}
\def\Dpm     {\ensuremath{\D^\pm}\xspace}
\def\Dmp     {\ensuremath{\D^\mp}\xspace}
\def\DpDm    {\ensuremath{\Dp {\kern -0.16em \Dm}}\xspace}
\def\Dstar   {\ensuremath{\D^*}\xspace}
\def\Dstarb  {\ensuremath{\Dbar^*}\xspace}
\def\Dstarz  {\ensuremath{\D^{*0}}\xspace}
\def\Dstarzb {\ensuremath{\Dbar^{*0}}\xspace}
\def\Dstarp  {\ensuremath{\D^{*+}}\xspace}
\def\Dstarm  {\ensuremath{\D^{*-}}\xspace}
\def\Dstarpm {\ensuremath{\D^{*\pm}}\xspace}
\def\Dstarmp {\ensuremath{\D^{*\mp}}\xspace}
\def\Ds      {\ensuremath{\D^+_\squark}\xspace}
\def\Dsp     {\ensuremath{\D^+_\squark}\xspace}
\def\Dsm     {\ensuremath{\D^-_\squark}\xspace}
\def\Dspm    {\ensuremath{\D^{\pm}_\squark}\xspace}
\def\Dss     {\ensuremath{\D^{*+}_\squark}\xspace}
\def\Dssp    {\ensuremath{\D^{*+}_\squark}\xspace}
\def\Dssm    {\ensuremath{\D^{*-}_\squark}\xspace}
\def\Dsspm   {\ensuremath{\D^{*\pm}_\squark}\xspace}

\def\B       {\ensuremath{\PB}\xspace}
  \def\Bbar    {\kern 0.18em\overline{\kern -0.18em \PB}{}\xspace}
\def\Bb      {\ensuremath{\Bbar}\xspace}
\def\BBbar   {\ensuremath{\B\Bbar}\xspace} 
\def\Bz      {\ensuremath{\B^0}\xspace}
\def\Bzb     {\ensuremath{\Bbar^0}\xspace}
\def\Bu      {\ensuremath{\B^+}\xspace}
\def\Bub     {\ensuremath{\B^-}\xspace}
\def\Bp      {\ensuremath{\Bu}\xspace}
\def\Bm      {\ensuremath{\Bub}\xspace}
\def\Bpm     {\ensuremath{\B^\pm}\xspace}
\def\Bmp     {\ensuremath{\B^\mp}\xspace}
\def\Bd      {\ensuremath{\B^0}\xspace}
\def\Bs      {\ensuremath{\B^0_\squark}\xspace}
\def\Bsb     {\ensuremath{\Bbar^0_\squark}\xspace}
\def\Bstar   {\ensuremath{B_s^*}\xspace}
\def\Bstarb  {\ensuremath{\Bb_s^*}\xspace}
\def\Bdb     {\ensuremath{\Bbar^0}\xspace}
\def\Bc      {\ensuremath{\B_\cquark^+}\xspace}
\def\Bcp     {\ensuremath{\B_\cquark^+}\xspace}
\def\Bcm     {\ensuremath{\B_\cquark^-}\xspace}
\def\Bcpm    {\ensuremath{\B_\cquark^\pm}\xspace}


\def\jpsi     {\ensuremath{{\PJ\mskip -3mu/\mskip -2mu\Ppsi\mskip 2mu}}\xspace}
\def\psitwos  {\ensuremath{\Ppsi{(2S)}}\xspace}
\def\psiprpr  {\ensuremath{\Ppsi(3770)}\xspace}
\def\etac     {\ensuremath{\Peta_\cquark}\xspace}
\def\chiczero {\ensuremath{\Pchi_{\cquark 0}}\xspace}
\def\chicone  {\ensuremath{\Pchi_{\cquark 1}}\xspace}
\def\chictwo  {\ensuremath{\Pchi_{\cquark 2}}\xspace}
  \def\Y#1S{\ensuremath{\PUpsilon{(#1S)}}\xspace}
\def\OneS  {\Y1S}
\def\TwoS  {\Y2S}
\def\ThreeS{\Y3S}
\def\FourS {\Y4S}
\def\FiveS {\Y5S}

\def\chic  {\ensuremath{\Pchi_{c}}\xspace}


\def\proton      {\ensuremath{\Pp}\xspace}
\def\antiproton  {\ensuremath{\overline \proton}\xspace}
\def\neutron     {\ensuremath{\Pn}\xspace}
\def\antineutron {\ensuremath{\overline \neutron}\xspace}

\def\Deltares {\ensuremath{\PDelta}\xspace}
\def\Deltaresbar{\ensuremath{\overline \Deltares}\xspace}
\def\Xires {\ensuremath{\PXi}\xspace}
\def\Xiresbar{\ensuremath{\overline \Xires}\xspace}
\def\L {\ensuremath{\PLambda}\xspace}
\def\Lbar {\ensuremath{\kern 0.1em\overline{\kern -0.1em\Lambda\kern -0.05em}\kern 0.05em{}}\xspace}
\def\Lambdares {\ensuremath{\PLambda}\xspace}
\def\Lambdaresbar{\ensuremath{\Lbar}\xspace}
\def\Sigmares {\ensuremath{\PSigma}\xspace}
\def\Sigmaresbar{\ensuremath{\overline \Sigmares}\xspace}
\def\Omegares {\ensuremath{\POmega}\xspace}
\def\Omegaresbar{\ensuremath{\overline \Omegares}\xspace}


\def\Lb      {\ensuremath{\L^0_\bquark}\xspace}
\def\Lbbar   {\ensuremath{\Lbar^0_\bquark}\xspace}
\def\Lc      {\ensuremath{\L^+_\cquark}\xspace}
\def\Lcbar   {\ensuremath{\Lbar^-_\cquark}\xspace}


\def\BF         {{\ensuremath{\cal B}\xspace}}
\def\BRvis      {{\ensuremath{\BR_{\rm{vis}}}}}
\def\BR         {\BF}
\newcommand{\decay}[2]{\ensuremath{#1\!\to #2}\xspace}         
\def\ra                 {\ensuremath{\rightarrow}\xspace}
\def\to                 {\ensuremath{\rightarrow}\xspace}

\newcommand{\tauBs}{\ensuremath{\tau_{\Bs}}\xspace}
\newcommand{\tauBd}{\ensuremath{\tau_{\Bd}}\xspace}
\newcommand{\tauBz}{\ensuremath{\tau_{\Bz}}\xspace}
\newcommand{\tauBu}{\ensuremath{\tau_{\Bp}}\xspace}
\newcommand{\tauDp}{\ensuremath{\tau_{\Dp}}\xspace}
\newcommand{\tauDz}{\ensuremath{\tau_{\Dz}}\xspace}
\newcommand{\tauL}{\ensuremath{\tau_{\rm L}}\xspace}
\newcommand{\tauH}{\ensuremath{\tau_{\rm H}}\xspace}

\newcommand{\mBd}{\ensuremath{m_{\Bd}}\xspace}
\newcommand{\mBp}{\ensuremath{m_{\Bp}}\xspace}
\newcommand{\mBs}{\ensuremath{m_{\Bs}}\xspace}
\newcommand{\mBc}{\ensuremath{m_{\Bc}}\xspace}
\newcommand{\mLb}{\ensuremath{m_{\Lb}}\xspace}

\def\grpsuthree {\ensuremath{\mathrm{SU}(3)}\xspace}
\def\grpsutw    {\ensuremath{\mathrm{SU}(2)}\xspace}
\def\grpuone    {\ensuremath{\mathrm{U}(1)}\xspace}

\def\ssqtw {\ensuremath{\sin^{2}\!\theta_{\mathrm{W}}}\xspace}
\def\csqtw {\ensuremath{\cos^{2}\!\theta_{\mathrm{W}}}\xspace}
\def\stw   {\ensuremath{\sin\theta_{\mathrm{W}}}\xspace}
\def\ctw   {\ensuremath{\cos\theta_{\mathrm{W}}}\xspace}
\def\ssqtwef {\ensuremath{{\sin}^{2}\theta_{\mathrm{W}}^{\mathrm{eff}}}\xspace}
\def\csqtwef {\ensuremath{{\cos}^{2}\theta_{\mathrm{W}}^{\mathrm{eff}}}\xspace}
\def\stwef {\ensuremath{\sin\theta_{\mathrm{W}}^{\mathrm{eff}}}\xspace}
\def\ctwef {\ensuremath{\cos\theta_{\mathrm{W}}^{\mathrm{eff}}}\xspace}
\def\gv    {\ensuremath{g_{\mbox{\tiny V}}}\xspace}
\def\ga    {\ensuremath{g_{\mbox{\tiny A}}}\xspace}

\def\order   {\ensuremath{\mathcal{O}}\xspace}
\def\ordalph {\ensuremath{\mathcal{O}(\alpha)}\xspace}
\def\ordalsq {\ensuremath{\mathcal{O}(\alpha^{2})}\xspace}
\def\ordalcb {\ensuremath{\mathcal{O}(\alpha^{3})}\xspace}

\newcommand{\as}{\ensuremath{\alpha_{\scriptscriptstyle S}}\xspace}
\newcommand{\MSb}{\ensuremath{\overline{\mathrm{MS}}}\xspace}
\newcommand{\lqcd}{\ensuremath{\Lambda_{\mathrm{QCD}}}\xspace}
\def\qsq       {\ensuremath{q^2}\xspace}


\def\eps   {\ensuremath{\varepsilon}\xspace}
\def\epsK  {\ensuremath{\varepsilon_K}\xspace}
\def\epsB  {\ensuremath{\varepsilon_B}\xspace}
\def\epsp  {\ensuremath{\varepsilon^\prime_K}\xspace}

\def\CP                {\ensuremath{C\!P}\xspace}
\def\CPT               {\ensuremath{C\!PT}\xspace}

\def\rhobar {\ensuremath{\overline \rho}\xspace}
\def\etabar {\ensuremath{\overline \eta}\xspace}

\def\Vud  {\ensuremath{|V_{\uquark\dquark}|}\xspace}
\def\Vcd  {\ensuremath{|V_{\cquark\dquark}|}\xspace}
\def\Vtd  {\ensuremath{|V_{\tquark\dquark}|}\xspace}
\def\Vus  {\ensuremath{|V_{\uquark\squark}|}\xspace}
\def\Vcs  {\ensuremath{|V_{\cquark\squark}|}\xspace}
\def\Vts  {\ensuremath{|V_{\tquark\squark}|}\xspace}
\def\Vub  {\ensuremath{|V_{\uquark\bquark}|}\xspace}
\def\Vcb  {\ensuremath{|V_{\cquark\bquark}|}\xspace}
\def\Vtb  {\ensuremath{|V_{\tquark\bquark}|}\xspace}


\newcommand{\dm}{\ensuremath{\Delta m}\xspace}
\newcommand{\dms}{\ensuremath{\Delta m_{\squark}}\xspace}
\newcommand{\dmd}{\ensuremath{\Delta m_{\dquark}}\xspace}
\newcommand{\DG}{\ensuremath{\Delta\Gamma}\xspace}
\newcommand{\DGs}{\ensuremath{\Delta\Gamma_{\squark}}\xspace}
\newcommand{\DGd}{\ensuremath{\Delta\Gamma_{\dquark}}\xspace}
\newcommand{\Gs}{\ensuremath{\Gamma_{\squark}}\xspace}
\newcommand{\Gd}{\ensuremath{\Gamma_{\dquark}}\xspace}

\newcommand{\MBq}{\ensuremath{M_{\B_\quark}}\xspace}
\newcommand{\DGq}{\ensuremath{\Delta\Gamma_{\quark}}\xspace}
\newcommand{\Gq}{\ensuremath{\Gamma_{\quark}}\xspace}
\newcommand{\dmq}{\ensuremath{\Delta m_{\quark}}\xspace}
\newcommand{\GL}{\ensuremath{\Gamma_{\rm L}}\xspace}
\newcommand{\GH}{\ensuremath{\Gamma_{\rm H}}\xspace}

\newcommand{\DGsGs}{\ensuremath{\Delta\Gamma_{\squark}/\Gamma_{\squark}}\xspace}
\newcommand{\Delm}{\mbox{$\Delta m $}\xspace}
\newcommand{\ACP}{\ensuremath{{\cal A}^{\CP}}\xspace}
\newcommand{\Adir}{\ensuremath{{\cal A}^{\rm dir}}\xspace}
\newcommand{\Amix}{\ensuremath{{\cal A}^{\rm mix}}\xspace}
\newcommand{\ADelta}{\ensuremath{{\cal A}^\Delta}\xspace}
\newcommand{\phid}{\ensuremath{\phi_{\dquark}}\xspace}
\newcommand{\sinphid}{\ensuremath{\sin\!\phid}\xspace}
\newcommand{\phis}{\ensuremath{\phi_{\squark}}\xspace}
\newcommand{\betas}{\ensuremath{\beta_{\squark}}\xspace}
\newcommand{\sbetas}{\ensuremath{\sigma(\beta_{\squark})}\xspace}
\newcommand{\stbetas}{\ensuremath{\sigma(2\beta_{\squark})}\xspace}
\newcommand{\stphis}{\ensuremath{\sigma(\phi_{\squark})}\xspace}
\newcommand{\sinphis}{\ensuremath{\sin\!\phis}\xspace}

\newcommand{\edet}{{\ensuremath{\varepsilon_{\rm det}}}\xspace}
\newcommand{\erec}{{\ensuremath{\varepsilon_{\rm rec/det}}}\xspace}
\newcommand{\esel}{{\ensuremath{\varepsilon_{\rm sel/rec}}}\xspace}
\newcommand{\etrg}{{\ensuremath{\varepsilon_{\rm trg/sel}}}\xspace}
\newcommand{\etot}{{\ensuremath{\varepsilon_{\rm tot}}}\xspace}

\newcommand{\mistag}{\ensuremath{\omega}\xspace}
\newcommand{\wcomb}{\ensuremath{\omega^{\rm comb}}\xspace}
\newcommand{\etag}{{\ensuremath{\varepsilon_{\rm tag}}}\xspace}
\newcommand{\etagcomb}{{\ensuremath{\varepsilon_{\rm tag}^{\rm comb}}}\xspace}
\newcommand{\effeff}{\ensuremath{\varepsilon_{\rm eff}}\xspace}
\newcommand{\effeffcomb}{\ensuremath{\varepsilon_{\rm eff}^{\rm comb}}\xspace}
\newcommand{\efftag}{{\ensuremath{\etag(1-2\omega)^2}}\xspace}
\newcommand{\effD}{{\ensuremath{\etag D^2}}\xspace}

\newcommand{\etagprompt}{{\ensuremath{\varepsilon_{\rm tag}^{\rm Pr}}}\xspace}
\newcommand{\etagLL}{{\ensuremath{\varepsilon_{\rm tag}^{\rm LL}}}\xspace}


\def\BdToKstmm    {\decay{\Bd}{\Kstarz\mup\mun}}
\def\BdbToKstmm   {\decay{\Bdb}{\Kstarzb\mup\mun}}

\def\BsToJPsiPhi  {\decay{\Bs}{\jpsi\phi}}
\def\BdToJPsiKst  {\decay{\Bd}{\jpsi\Kstarz}}
\def\BdbToJPsiKst {\decay{\Bdb}{\jpsi\Kstarzb}}

\def\BsPhiGam     {\decay{\Bs}{\phi \g}}
\def\BdKstGam     {\decay{\Bd}{\Kstarz \g}}

\def\BTohh        {\decay{\B}{\Ph^+ \Ph'^-}}
\def\BdTopipi     {\decay{\Bd}{\pip\pim}}
\def\BdToKpi      {\decay{\Bd}{\Kp\pim}}
\def\BsToKK       {\decay{\Bs}{\Kp\Km}}
\def\BsTopiK      {\decay{\Bs}{\pip\Km}}

\def\BdKstee  {\decay{\Bd}{\Kstarz\epem}}
\def\BdbKstee {\decay{\Bdb}{\Kstarzb\epem}}
\def\bsll     {\decay{\bquark}{\squark \ell^+ \ell^-}}
\def\AFB      {\ensuremath{A_{\mathrm{FB}}}\xspace}
\def\FL       {\ensuremath{F_{\mathrm{L}}}\xspace}
\def\AT#1     {\ensuremath{A_{\mathrm{T}}^{#1}}\xspace}           
\def\btosgam  {\decay{\bquark}{\squark \g}}
\def\btodgam  {\decay{\bquark}{\dquark \g}}
\def\Bsmm     {\decay{\Bs}{\mup\mun}}
\def\Bdmm     {\decay{\Bd}{\mup\mun}}
\def\ctl       {\ensuremath{\cos{\theta_l}}\xspace}
\def\ctk       {\ensuremath{\cos{\theta_K}}\xspace}

\def\C#1      {\ensuremath{\mathcal{C}_{#1}}\xspace}                       
\def\Cp#1     {\ensuremath{\mathcal{C}_{#1}^{'}}\xspace}                    
\def\Ceff#1   {\ensuremath{\mathcal{C}_{#1}^{\mathrm{(eff)}}}\xspace}        
\def\Cpeff#1  {\ensuremath{\mathcal{C}_{#1}^{'\mathrm{(eff)}}}\xspace}       
\def\Ope#1    {\ensuremath{\mathcal{O}_{#1}}\xspace}                       
\def\Opep#1   {\ensuremath{\mathcal{O}_{#1}^{'}}\xspace}                    


\def\xprime     {\ensuremath{x^{\prime}}\xspace}
\def\yprime     {\ensuremath{y^{\prime}}\xspace}
\def\ycp        {\ensuremath{y_{\CP}}\xspace}
\def\agamma     {\ensuremath{A_{\Gamma}}\xspace}
\def\kpi        {\ensuremath{\PK\Ppi}\xspace}
\def\kk         {\ensuremath{\PK\PK}\xspace}
\def\dkpi       {\decay{\PD}{\PK\Ppi}}
\def\dkk        {\decay{\PD}{\PK\PK}}
\def\dkpicf     {\decay{\Dz}{\Km\pip}}

\newcommand{\bra}[1]{\ensuremath{\langle #1|}}             
\newcommand{\ket}[1]{\ensuremath{|#1\rangle}}              
\newcommand{\braket}[2]{\ensuremath{\langle #1|#2\rangle}} 

\newcommand{\unit}[1]{\ensuremath{\rm\,#1}\xspace}          

\newcommand{\tev}{\ensuremath{\mathrm{\,Te\kern -0.1em V}}\xspace}
\newcommand{\gev}{\ensuremath{\mathrm{\,Ge\kern -0.1em V}}\xspace}
\newcommand{\mev}{\ensuremath{\mathrm{\,Me\kern -0.1em V}}\xspace}
\newcommand{\kev}{\ensuremath{\mathrm{\,ke\kern -0.1em V}}\xspace}
\newcommand{\ev}{\ensuremath{\mathrm{\,e\kern -0.1em V}}\xspace}
\newcommand{\gevc}{\ensuremath{{\mathrm{\,Ge\kern -0.1em V\!/}c}}\xspace}
\newcommand{\mevc}{\ensuremath{{\mathrm{\,Me\kern -0.1em V\!/}c}}\xspace}
\newcommand{\gevcc}{\ensuremath{{\mathrm{\,Ge\kern -0.1em V\!/}c^2}}\xspace}
\newcommand{\gevgevcccc}{\ensuremath{{\mathrm{\,Ge\kern -0.1em V^2\!/}c^4}}\xspace}
\newcommand{\mevcc}{\ensuremath{{\mathrm{\,Me\kern -0.1em V\!/}c^2}}\xspace}

\def\km   {\ensuremath{\rm \,km}\xspace}
\def\m    {\ensuremath{\rm \,m}\xspace}
\def\cm   {\ensuremath{\rm \,cm}\xspace}
\def\cma  {\ensuremath{{\rm \,cm}^2}\xspace}
\def\mm   {\ensuremath{\rm \,mm}\xspace}
\def\mma  {\ensuremath{{\rm \,mm}^2}\xspace}
\def\mum  {\ensuremath{\,\upmu\rm m}\xspace}
\def\muma {\ensuremath{\,\upmu\rm m^2}\xspace}
\def\nm   {\ensuremath{\rm \,nm}\xspace}
\def\fm   {\ensuremath{\rm \,fm}\xspace}
\def\barn{\ensuremath{\rm \,b}\xspace}
\def\barnhyph{\ensuremath{\rm -b}\xspace}
\def\mbarn{\ensuremath{\rm \,mb}\xspace}
\def\mub{\ensuremath{\rm \,\upmu b}\xspace}
\def\mbarnhyph{\ensuremath{\rm -mb}\xspace}
\def\nb {\ensuremath{\rm \,nb}\xspace}
\def\invnb {\ensuremath{\mbox{\,nb}^{-1}}\xspace}
\def\pb {\ensuremath{\rm \,pb}\xspace}
\def\invpb {\ensuremath{\mbox{\,pb}^{-1}}\xspace}
\def\fb   {\ensuremath{\mbox{\,fb}}\xspace}
\def\invfb   {\ensuremath{\mbox{\,fb}^{-1}}\xspace}

\def\sec  {\ensuremath{\rm {\,s}}\xspace}
\def\ms   {\ensuremath{{\rm \,ms}}\xspace}
\def\mus  {\ensuremath{\,\upmu{\rm s}}\xspace}
\def\ns   {\ensuremath{{\rm \,ns}}\xspace}
\def\ps   {\ensuremath{{\rm \,ps}}\xspace}
\def\fs   {\ensuremath{\rm \,fs}\xspace}

\def\mhz  {\ensuremath{{\rm \,MHz}}\xspace}
\def\khz  {\ensuremath{{\rm \,kHz}}\xspace}
\def\hz   {\ensuremath{{\rm \,Hz}}\xspace}

\def\invps{\ensuremath{{\rm \,ps^{-1}}}\xspace}

\def\yr   {\ensuremath{\rm \,yr}\xspace}
\def\hr   {\ensuremath{\rm \,hr}\xspace}

\def\degc {\ensuremath{^\circ}{C}\xspace}
\def\degk {\ensuremath {\rm K}\xspace}

\def\Xrad {\ensuremath{X_0}\xspace}
\def\NIL{\ensuremath{\lambda_{int}}\xspace}
\def\mip {MIP\xspace}
\def\neutroneq {\ensuremath{\rm \,n_{eq}}\xspace}
\def\neqcmcm {\ensuremath{\rm \,n_{eq} / cm^2}\xspace}
\def\kRad {\ensuremath{\rm \,kRad}\xspace}
\def\MRad {\ensuremath{\rm \,MRad}\xspace}
\def\ci {\ensuremath{\rm \,Ci}\xspace}
\def\mci {\ensuremath{\rm \,mCi}\xspace}

\def\sx    {\ensuremath{\sigma_x}\xspace}    
\def\sy    {\ensuremath{\sigma_y}\xspace}   
\def\sz    {\ensuremath{\sigma_z}\xspace}    

\newcommand{\stat}{\ensuremath{\mathrm{(stat)}}\xspace}
\newcommand{\syst}{\ensuremath{\mathrm{(syst)}}\xspace}


\def\order{{\ensuremath{\cal O}}\xspace}
\newcommand{\chisq}{\ensuremath{\chi^2}\xspace}

\def\deriv {\ensuremath{\mathrm{d}}}

\def\gsim{{~\raise.15em\hbox{$>$}\kern-.85em
          \lower.35em\hbox{$\sim$}~}\xspace}
\def\lsim{{~\raise.15em\hbox{$<$}\kern-.85em
          \lower.35em\hbox{$\sim$}~}\xspace}

\newcommand{\mean}[1]{\ensuremath{\left\langle #1 \right\rangle}} 
\newcommand{\abs}[1]{\ensuremath{\left\|#1\right\|}} 
\newcommand{\Real}{\ensuremath{\mathcal{R}e}\xspace}
\newcommand{\Imag}{\ensuremath{\mathcal{I}m}\xspace}

\def\PDF {PDF\xspace}

\def\Ebeam {\ensuremath{E_{\mbox{\tiny BEAM}}}\xspace}
\def\sqs   {\ensuremath{\protect\sqrt{s}}\xspace}

\def\ptot       {\mbox{$p$}\xspace}
\def\pt         {\mbox{$p_{\rm T}$}\xspace}
\def\et         {\mbox{$E_{\rm T}$}\xspace}
\def\dpp        {\ensuremath{\mathrm{d}\hspace{-0.1em}p/p}\xspace}

\newcommand{\dedx}{\ensuremath{\mathrm{d}\hspace{-0.1em}E/\mathrm{d}x}\xspace}


\def\dllkpi     {\ensuremath{\mathrm{DLL}_{\kaon\pion}}\xspace}
\def\dllppi     {\ensuremath{\mathrm{DLL}_{\proton\pion}}\xspace}
\def\dllepi     {\ensuremath{\mathrm{DLL}_{\electron\pion}}\xspace}
\def\dllmupi    {\ensuremath{\mathrm{DLL}_{\mmu\pi}}\xspace}

\def\mphi       {\mbox{$\phi$}\xspace}
\def\mtheta     {\mbox{$\theta$}\xspace}
\def\ctheta     {\mbox{$\cos\theta$}\xspace}
\def\stheta     {\mbox{$\sin\theta$}\xspace}
\def\ttheta     {\mbox{$\tan\theta$}\xspace}

\def\degrees{\ensuremath{^{\circ}}\xspace}
\def\krad {\ensuremath{\rm \,krad}\xspace}
\def\mrad{\ensuremath{\rm \,mrad}\xspace}
\def\rad{\ensuremath{\rm \,rad}\xspace}

\def\betastar {\ensuremath{\beta^*}}
\newcommand{\lum} {\ensuremath{\mathcal{L}}\xspace}
\newcommand{\intlum}[1]{\ensuremath{\int\lum=#1\xspace}}  


\def\evtgen     {\mbox{\textsc{EvtGen}}\xspace}
\def\pythia     {\mbox{\textsc{Pythia}}\xspace}
\def\fluka      {\mbox{\textsc{Fluka}}\xspace}
\def\tosca      {\mbox{\textsc{Tosca}}\xspace}
\def\ansys      {\mbox{\textsc{Ansys}}\xspace}
\def\spice      {\mbox{\textsc{Spice}}\xspace}
\def\garfield   {\mbox{\textsc{Garfield}}\xspace}
\def\geant      {\mbox{\textsc{Geant3}}\xspace}
\def\hepmc      {\mbox{\textsc{HepMC}}\xspace}
\def\gauss      {\mbox{\textsc{Gauss}}\xspace}
\def\gaudi      {\mbox{\textsc{Gaudi}}\xspace}
\def\boole      {\mbox{\textsc{Boole}}\xspace}
\def\brunel     {\mbox{\textsc{Brunel}}\xspace}
\def\davinci    {\mbox{\textsc{DaVinci}}\xspace}
\def\erasmus    {\mbox{\textsc{Erasmus}}\xspace}
\def\moore      {\mbox{\textsc{Moore}}\xspace}
\def\ganga      {\mbox{\textsc{Ganga}}\xspace}
\def\dirac      {\mbox{\textsc{Dirac}}\xspace}
\def\root       {\mbox{\textsc{Root}}\xspace}
\def\roofit     {\mbox{\textsc{RooFit}}\xspace}
\def\pyroot     {\mbox{\textsc{PyRoot}}\xspace}
\def\photos     {\mbox{\textsc{Photos}}\xspace}

\def\cpp        {\mbox{\textsc{C\raisebox{0.1em}{{\footnotesize{++}}}}}\xspace}
\def\python     {\mbox{\textsc{Python}}\xspace}
\def\ruby       {\mbox{\textsc{Ruby}}\xspace}
\def\fortran    {\mbox{\textsc{Fortran}}\xspace}
\def\svn        {\mbox{\textsc{SVN}}\xspace}

\def\kbytes     {\ensuremath{{\rm \,kbytes}}\xspace}
\def\kbsps      {\ensuremath{{\rm \,kbytes/s}}\xspace}
\def\kbits      {\ensuremath{{\rm \,kbits}}\xspace}
\def\kbsps      {\ensuremath{{\rm \,kbits/s}}\xspace}
\def\mbsps      {\ensuremath{{\rm \,Mbits/s}}\xspace}
\def\mbytes     {\ensuremath{{\rm \,Mbytes}}\xspace}
\def\mbps       {\ensuremath{{\rm \,Mbyte/s}}\xspace}
\def\mbsps      {\ensuremath{{\rm \,Mbytes/s}}\xspace}
\def\gbsps      {\ensuremath{{\rm \,Gbits/s}}\xspace}
\def\gbytes     {\ensuremath{{\rm \,Gbytes}}\xspace}
\def\gbsps      {\ensuremath{{\rm \,Gbytes/s}}\xspace}
\def\tbytes     {\ensuremath{{\rm \,Tbytes}}\xspace}
\def\tbpy       {\ensuremath{{\rm \,Tbytes/yr}}\xspace}

\def\dst        {DST\xspace}


\def\nonn {\ensuremath{\rm {\it{n^+}}\mbox{-}on\mbox{-}{\it{n}}}\xspace}
\def\ponn {\ensuremath{\rm {\it{p^+}}\mbox{-}on\mbox{-}{\it{n}}}\xspace}
\def\nonp {\ensuremath{\rm {\it{n^+}}\mbox{-}on\mbox{-}{\it{p}}}\xspace}
\def\cvd  {CVD\xspace}
\def\mwpc {MWPC\xspace}
\def\gem  {GEM\xspace}

\def\tell1  {TELL1\xspace}
\def\ukl1   {UKL1\xspace}
\def\beetle {Beetle\xspace}
\def\otis   {OTIS\xspace}
\def\croc   {CROC\xspace}
\def\carioca {CARIOCA\xspace}
\def\dialog {DIALOG\xspace}
\def\sync   {SYNC\xspace}
\def\cardiac {CARDIAC\xspace}
\def\gol    {GOL\xspace}
\def\vcsel  {VCSEL\xspace}
\def\ttc    {TTC\xspace}
\def\ttcrx  {TTCrx\xspace}
\def\hpd    {HPD\xspace}
\def\pmt    {PMT\xspace}
\def\specs  {SPECS\xspace}
\def\elmb   {ELMB\xspace}
\def\fpga   {FPGA\xspace}
\def\plc    {PLC\xspace}
\def\rasnik {RASNIK\xspace}
\def\elmb   {ELMB\xspace}
\def\can    {CAN\xspace}
\def\lvds   {LVDS\xspace}
\def\ntc    {NTC\xspace}
\def\adc    {ADC\xspace}
\def\led    {LED\xspace}
\def\ccd    {CCD\xspace}
\def\hv     {HV\xspace}
\def\lv     {LV\xspace}
\def\pvss   {PVSS\xspace}
\def\cmos   {CMOS\xspace}
\def\fifo   {FIFO\xspace}
\def\ccpc   {CCPC\xspace}

\def\cfourften     {\ensuremath{\rm C_4 F_{10}}\xspace}
\def\cffour        {\ensuremath{\rm CF_4}\xspace}
\def\cotwo         {\ensuremath{\rm CO_2}\xspace} 
\def\csixffouteen  {\ensuremath{\rm C_6 F_{14}}\xspace} 
\def\mgftwo     {\ensuremath{\rm Mg F_2}\xspace} 
\def\siotwo     {\ensuremath{\rm SiO_2}\xspace} 

\newcommand{\eg}{\mbox{\itshape e.g.}\xspace}
\newcommand{\ie}{\mbox{\itshape i.e.}}
\newcommand{\etal}{{\slshape et al.\/}\xspace}
\newcommand{\etc}{\mbox{\itshape etc.}\xspace}
\newcommand{\cf}{\mbox{\itshape cf.}\xspace}
\newcommand{\ffp}{\mbox{\itshape ff.}\xspace}

\def\Kbar    {\ensuremath{\overline{K}{}^0}\xspace}
\def\KS      {\ensuremath{K^0_S}\xspace} 
\def\KSb     {\ensuremath{\Kbar^0_S}\xspace} 
\def\KL      {\ensuremath{K^0_L}\xspace} 








\title{ 
Measurement of the branching fraction and time-dependent  \mbox{\boldmath\CP} asymmetry
for \mbox{\boldmath$\Bz\to\jpsi\piz$} decays }




\noaffiliation
\affiliation{University of the Basque Country UPV/EHU, 48080 Bilbao}
\affiliation{Beihang University, Beijing 100191}
\affiliation{Brookhaven National Laboratory, Upton, New York 11973}
\affiliation{Budker Institute of Nuclear Physics SB RAS, Novosibirsk 630090}
\affiliation{Faculty of Mathematics and Physics, Charles University, 121 16 Prague}
\affiliation{University of Cincinnati, Cincinnati, Ohio 45221}
\affiliation{Deutsches Elektronen--Synchrotron, 22607 Hamburg}
\affiliation{University of Florida, Gainesville, Florida 32611}
\affiliation{Key Laboratory of Nuclear Physics and Ion-beam Application (MOE) and Institute of Modern Physics, Fudan University, Shanghai 200443}
\affiliation{Gifu University, Gifu 501-1193}
\affiliation{II. Physikalisches Institut, Georg-August-Universit\"at G\"ottingen, 37073 G\"ottingen}
\affiliation{SOKENDAI (The Graduate University for Advanced Studies), Hayama 240-0193}
\affiliation{Hanyang University, Seoul 133-791}
\affiliation{University of Hawaii, Honolulu, Hawaii 96822}
\affiliation{High Energy Accelerator Research Organization (KEK), Tsukuba 305-0801}
\affiliation{J-PARC Branch, KEK Theory Center, High Energy Accelerator Research Organization (KEK), Tsukuba 305-0801}
\affiliation{Forschungszentrum J\"{u}lich, 52425 J\"{u}lich}
\affiliation{IKERBASQUE, Basque Foundation for Science, 48013 Bilbao}
\affiliation{Indian Institute of Technology Bhubaneswar, Satya Nagar 751007}
\affiliation{Indian Institute of Technology Guwahati, Assam 781039}
\affiliation{Indian Institute of Technology Hyderabad, Telangana 502285}
\affiliation{Indian Institute of Technology Madras, Chennai 600036}
\affiliation{Indiana University, Bloomington, Indiana 47408}
\affiliation{Institute of High Energy Physics, Chinese Academy of Sciences, Beijing 100049}
\affiliation{Institute of High Energy Physics, Vienna 1050}
\affiliation{Institute for High Energy Physics, Protvino 142281}
\affiliation{INFN - Sezione di Napoli, 80126 Napoli}
\affiliation{INFN - Sezione di Torino, 10125 Torino}
\affiliation{Advanced Science Research Center, Japan Atomic Energy Agency, Naka 319-1195}
\affiliation{J. Stefan Institute, 1000 Ljubljana}
\affiliation{Institut f\"ur Experimentelle Teilchenphysik, Karlsruher Institut f\"ur Technologie, 76131 Karlsruhe}
\affiliation{Kavli Institute for the Physics and Mathematics of the Universe (WPI), University of Tokyo, Kashiwa 277-8583}
\affiliation{Kennesaw State University, Kennesaw, Georgia 30144}
\affiliation{King Abdulaziz City for Science and Technology, Riyadh 11442}
\affiliation{Department of Physics, Faculty of Science, King Abdulaziz University, Jeddah 21589}
\affiliation{Korea Institute of Science and Technology Information, Daejeon 305-806}
\affiliation{Korea University, Seoul 136-713}
\affiliation{Kyungpook National University, Daegu 702-701}
\affiliation{LAL, Univ. Paris-Sud, CNRS/IN2P3, Universit\'{e} Paris-Saclay, Orsay}
\affiliation{\'Ecole Polytechnique F\'ed\'erale de Lausanne (EPFL), Lausanne 1015}
\affiliation{P.N. Lebedev Physical Institute of the Russian Academy of Sciences, Moscow 119991}
\affiliation{Faculty of Mathematics and Physics, University of Ljubljana, 1000 Ljubljana}
\affiliation{Ludwig Maximilians University, 80539 Munich}
\affiliation{Luther College, Decorah, Iowa 52101}
\affiliation{Malaviya National Institute of Technology Jaipur, Jaipur 302017}
\affiliation{University of Maribor, 2000 Maribor}
\affiliation{Max-Planck-Institut f\"ur Physik, 80805 M\"unchen}
\affiliation{School of Physics, University of Melbourne, Victoria 3010}
\affiliation{University of Mississippi, University, Mississippi 38677}
\affiliation{University of Miyazaki, Miyazaki 889-2192}
\affiliation{Moscow Physical Engineering Institute, Moscow 115409}
\affiliation{Moscow Institute of Physics and Technology, Moscow Region 141700}
\affiliation{Graduate School of Science, Nagoya University, Nagoya 464-8602}
\affiliation{Universit\`{a} di Napoli Federico II, 80055 Napoli}
\affiliation{Nara Women's University, Nara 630-8506}
\affiliation{National Central University, Chung-li 32054}
\affiliation{National United University, Miao Li 36003}
\affiliation{Department of Physics, National Taiwan University, Taipei 10617}
\affiliation{H. Niewodniczanski Institute of Nuclear Physics, Krakow 31-342}
\affiliation{Nippon Dental University, Niigata 951-8580}
\affiliation{Niigata University, Niigata 950-2181}
\affiliation{Novosibirsk State University, Novosibirsk 630090}
\affiliation{Osaka City University, Osaka 558-8585}
\affiliation{Pacific Northwest National Laboratory, Richland, Washington 99352}
\affiliation{Panjab University, Chandigarh 160014}
\affiliation{Peking University, Beijing 100871}
\affiliation{University of Pittsburgh, Pittsburgh, Pennsylvania 15260}
\affiliation{Punjab Agricultural University, Ludhiana 141004}
\affiliation{Theoretical Research Division, Nishina Center, RIKEN, Saitama 351-0198}
\affiliation{University of Science and Technology of China, Hefei 230026}
\affiliation{Showa Pharmaceutical University, Tokyo 194-8543}
\affiliation{Soongsil University, Seoul 156-743}
\affiliation{University of South Carolina, Columbia, South Carolina 29208}
\affiliation{Stefan Meyer Institute for Subatomic Physics, Vienna 1090}
\affiliation{Sungkyunkwan University, Suwon 440-746}
\affiliation{School of Physics, University of Sydney, New South Wales 2006}
\affiliation{Department of Physics, Faculty of Science, University of Tabuk, Tabuk 71451}
\affiliation{Tata Institute of Fundamental Research, Mumbai 400005}
\affiliation{Department of Physics, Technische Universit\"at M\"unchen, 85748 Garching}
\affiliation{Toho University, Funabashi 274-8510}
\affiliation{Department of Physics, Tohoku University, Sendai 980-8578}
\affiliation{Earthquake Research Institute, University of Tokyo, Tokyo 113-0032}
\affiliation{Department of Physics, University of Tokyo, Tokyo 113-0033}
\affiliation{Tokyo Institute of Technology, Tokyo 152-8550}
\affiliation{Tokyo Metropolitan University, Tokyo 192-0397}
\affiliation{Virginia Polytechnic Institute and State University, Blacksburg, Virginia 24061}
\affiliation{Wayne State University, Detroit, Michigan 48202}
\affiliation{Yamagata University, Yamagata 990-8560}
\affiliation{Yonsei University, Seoul 120-749}
 \author{B.~Pal}\affiliation{Brookhaven National Laboratory, Upton, New York 11973} \affiliation{University of Cincinnati, Cincinnati, Ohio 45221} 
  \author{A.~J.~Schwartz}\affiliation{University of Cincinnati, Cincinnati, Ohio 45221} 
  \author{H.~Aihara}\affiliation{Department of Physics, University of Tokyo, Tokyo 113-0033} 
  \author{S.~Al~Said}\affiliation{Department of Physics, Faculty of Science, University of Tabuk, Tabuk 71451}\affiliation{Department of Physics, Faculty of Science, King Abdulaziz University, Jeddah 21589} 
  \author{D.~M.~Asner}\affiliation{Brookhaven National Laboratory, Upton, New York 11973} 
  \author{H.~Atmacan}\affiliation{University of South Carolina, Columbia, South Carolina 29208} 
  \author{V.~Aulchenko}\affiliation{Budker Institute of Nuclear Physics SB RAS, Novosibirsk 630090}\affiliation{Novosibirsk State University, Novosibirsk 630090} 
  \author{T.~Aushev}\affiliation{Moscow Institute of Physics and Technology, Moscow Region 141700} 
  \author{R.~Ayad}\affiliation{Department of Physics, Faculty of Science, University of Tabuk, Tabuk 71451} 
  \author{I.~Badhrees}\affiliation{Department of Physics, Faculty of Science, University of Tabuk, Tabuk 71451}\affiliation{King Abdulaziz City for Science and Technology, Riyadh 11442} 
% ...
\collaboration{The Belle Collaboration}


\begin{abstract}
\noindent
We measure the branching fraction and time-dependent \CP-violating asymmetry
for $\Bz\to\jpsi\piz$ decays using a data sample 
of 711~\invfb collected on the \FourS resonance by the Belle experiment running 
at the KEKB \epem collider. The branching fraction is measured to be 
$\BF(\Bz\to\jpsi\piz) = [1.62\,\pm 0.11~\stat \pm0.06~\syst]\times 10^{-5}$, 
which is the most precise measurement to date. The measured \CP asymmetry
parameters are $\mathcal{S} = -0.59\,\pm 0.19~\stat \pm 0.03~\syst$ and 
$\mathcal{A} = -0.15\,\pm 0.14~\stat\,^{+0.04}_{-0.03}~\syst$.
The mixing-induced \CP asymmetry ($\mathcal{S}$) differs from the case
of no \CP violation by 3.0 standard deviations, and the direct \CP asymmetry
($\mathcal{A}$) is consistent with zero.
\end{abstract}


\maketitle
\begin{figure}[htb]
\centering
\includegraphics[height=0.85in,width=0.244\textwidth]{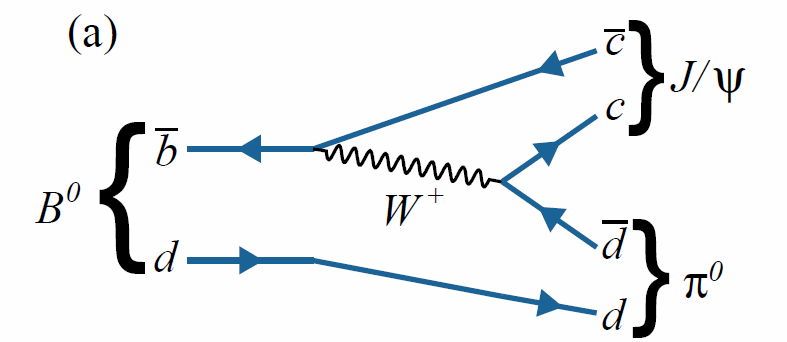}%
\includegraphics[height=0.85in,width=0.244\textwidth]{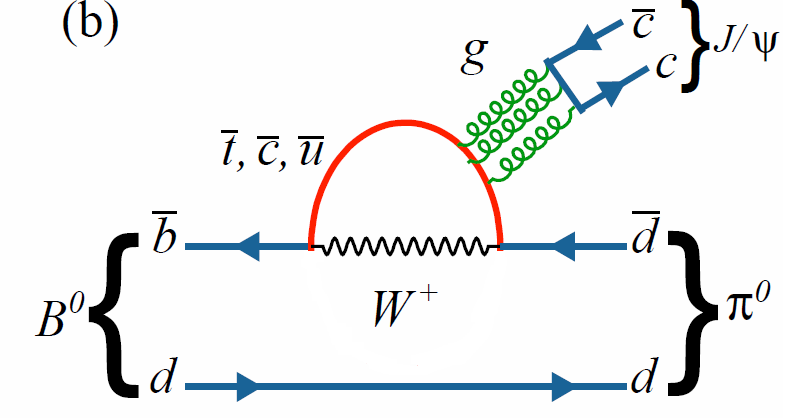}
\caption{\small (a) Tree and (b) penguin amplitudes for the decay $\Bz\to\jpsi\piz$.}
\label{fig:Feynman}
\end{figure}

\tighten
At  the quark level, the decay $\Bz\to\jpsi\piz$ proceeds via $b\to\ccbar d$ 
``tree'' and ``penguin'' amplitudes, as shown in Fig.~\ref{fig:Feynman}. Both 
amplitudes are suppressed in the Standard Model (the first one is color- and Cabibbo-suppressed), 
and thus the branching fraction
is small. The tree-level amplitude has the same weak phase as that of the 
$b\to\ccbar s$ amplitude governing, $e.g.$, $B^0\to \jpsi \KS$ decays,
while the penguin amplitude has a different weak phase. The former dominates mixing-induced \CP violation, while the addition of the latter gives rise to 
direct \CP violation.

In the process $\FourS\to\Bz\Bzb$, one of the two $B$ mesons can decay 
into a \CP eigenstate $f^{}_{\CP}$ at time $t_{\CP}$, while the other can decay 
into a flavor-specific state $f_{\rm tag}$ at time $t_{\rm tag}$.  The decay 
time evolution for the $B\to f^{}_{\CP}$  is~\cite{Carter:1980tk}
\begin{eqnarray}
\label{eq:one}
\mathcal{P}(\Delta t,q) & = & \frac{e^{-|\Delta t|/\tau_{\Bz}}}{4\tau_{\Bz}}\times \\ 
 & & \biggl( 1 + q\bigl[ \mathcal{S}\sin(\Delta m^{}_d\Delta t) 
+ \mathcal{A}\cos(\Delta m^{}_d\Delta t)  \bigr] \biggr),\nonumber
\end{eqnarray}
where 
$\Delta t=t_{\CP}-t_{\rm tag}$ is the difference in proper decay times between the 
two $B$ mesons; $q=+1~(-1)$ for  signal $\Bzb ~(\Bz)$ decays; 
$\Delta m^{}_d$ is the mass difference between the two mass eigenstates of 
the $\Bz$-$\Bzb$ system; and $\tau^{}_{B^0}$ is the $B^0$ lifetime.
The parameters $\mathcal{S}$ and $\mathcal{A}$ are \CP-violating and characterize 
mixing-induced and direct \CP violation, respectively. In the absence of the 
penguin amplitude, $\mathcal{A}=0$ and $\mathcal{S} = -\sin(2\phi_1)$, where
$\phi^{}_1 = {\rm arg}\bigl[ -(V^*_{cb}V^{}_{cd})/(V^*_{tb}V^{}_{td})\bigr]$.
However, this amplitude
 and any new physics (NP) process having a different weak phase  will shift 
$\mathcal{S}$ and $\mathcal{A}$ from these values. Thus, measuring these parameters
provides a way to search for NP. The values of $\mathcal{S}$ and $\mathcal{A}$ 
measured in $\Bz\to\jpsi\piz$ decays can also be used to constrain the small penguin 
contribution to $\Bz\to\jpsi\KS$ decays~\cite{Ciuchini:2005mg, Faller:2008zc, 
Jung:2012mp, DeBruyn:2014oga, Ligeti:2015yma, Frings:2015eva}. This small 
contribution is important as the decay $\Bz\to\jpsi\KS$ provides the 
most precise determination of $\phi^{}_1$.

The parameter $\mathcal{S}$ for $\Bz\to\jpsi\piz$ has previously been measured 
by Belle~\cite{Lee:2007wd} and BaBar~\cite{Aubert:2008bs}, but the results are 
not in good agreement. The BaBar result lies outside the physically allowed 
region, but the uncertainties are large. The previous result from Belle
was based on $535\times 10^6$ $\BBbar$ pairs~\cite{Lee:2007wd}. Here we 
update that measurement using the final Belle data set of $772\times 10^6$ $\BBbar$ 
pairs. We also update the $\Bz\to\jpsi\piz$ branching fraction, for which our previous 
measurement used only $32\times 10^6$ $\BBbar$ pairs~\cite{Abe:2002rc}. 
In addition to more data, the analysis presented here also uses improved tracking 
and photon reconstruction.


The Belle detector is a large-solid-angle magnetic spectrometer consisting
of a  silicon vertex detector (SVD), a 50-layer central
drift chamber (CDC), an array of aerogel threshold Cherenkov counters,
a barrel-like arrangement of time-of-flight scintillation counters,
and an electromagnetic calorimeter (ECL) comprising  CsI(Tl) crystals. 
These detector components are 
located inside a superconducting solenoid coil that provides a 1.5~T
magnetic field.  An iron flux-return (KLM) located outside the coil 
is instrumented to detect $K_L^0$ mesons and to identify muons.
Two inner detector configurations were used: a 2.0 cm radius beampipe
and a three-layer SVD were used for the first 
$152\times 10^6$ $\BBbar$ pairs of data, 
while a 1.5 cm radius beampipe, a four-layer SVD, and a 
small-cell inner drift chamber were used for the remaining
$620\times10^6$ $\BBbar$ pairs of data.
The detector is described in detail in Ref.~\cite{Belle}.
Event selection requirements are optimized using Monte Carlo (MC) simulation.
MC events are generated using \evtgen~\cite{Lange:2001uf}, and the detector 
response is modeled using \geant~\cite{geant3}.  Final-state radiation is 
taken into account using the \photos package~\cite{Golonka:2005pn}.  


The \FourS is produced with a Lorentz boost of $\beta\gamma=0.425$ along the $+z$ axis, 
which is defined as anti-parallel to the $\ep$ beam direction.  Since the $\Bz$ and 
$\Bzb$ mesons are approximately at rest in the \FourS center-of-mass (CM) system, 
$\Delta t$ is determined from the displacement in $z$ between the two $B$ decay 
vertices: $\Delta t \approx \Delta z/c\beta\gamma$.

The reconstruction of $\Bz\to\jpsi\piz$ proceeds by first reconstructing 
$\piz\to\gamma\gamma$ 
candidates. An ECL cluster not matched to any track is identified as a photon candidate. 
Such candidates are required to have an energy greater than 50~MeV in the barrel region 
and greater than 100~MeV in the end-cap regions, where the barrel region covers the polar 
angle $32^{\circ} < \theta < 130^{\circ}$ and the end-cap regions cover the ranges 
$12^{\circ} < \theta < 32^{\circ}$ and $130^{\circ}<\theta <157^{\circ}$. We require 
that the $\gamma\gamma$ invariant mass be within  $20 \mevcc$ (about $3.5\sigma$ 
in resolution) of the $\piz$ mass~\cite{PDG}. 
To improve the $\pi^0$ momentum resolution, 
we perform a mass-constrained fit and require that the resulting $\chi^2$ be less than~30. 
This requirement is relatively loose, retaining  more than 99\% of events.

We subsequently combine $\piz$ candidates with $\jpsi$ candidates, which are reconstructed 
in the $e^+e^-$ and $\mu^+\mu^-$ decay channels. All charged tracks are required to have a 
minimum number of SVD hits: $\ge 2$ in the beam direction, and $\ge 1$ in the transverse 
direction. Electron identification is based on the ratio of the ECL cluster energy to 
the particle momentum as measured in the CDC, as well as the position and shape of the 
electromagnetic shower in the ECL.
In order to account for radiative energy loss in $\epem$ decays, we include up 
to two bremsstrahlung photons that lie within 50~mrad of each of the reconstructed
tracks when calculating the $e^+$ and $e^-$ four-momenta. 
Muons are identified by corresponding hit positions and the track penetration 
depth in the KLM. The reconstructed $\jpsi$ invariant masses $M^{}_{ee(\g)}$ 
and $M^{}_{\mu\mu}$ are required to satisfy 
$-150 \mevcc <M^{}_{ee(\g)}-m^{}_{\jpsi}< +36 \mevcc$ and 
$-60 \mevcc <M^{}_{\mu\mu}-m^{}_{\jpsi}<+36 \mevcc$, where 
$m^{}_{\jpsi}$ is the nominal $\jpsi$ mass~\cite{PDG}.
The asymmetric mass ranges account for the radiative tail, which biases
the reconstructed mass towards lower values.  For selected $\jpsi$ candidates, 
vertex- and mass-constrained fits are performed to improve the momentum resolution. 

Candidate $\Bz$ mesons are identified using the beam-energy-constrained mass
$M_{\rm bc}= \Big(\sqrt{E^2_{\rm beam} - |{\vec{p}_B}|^2c^2}\Big)/c^2$, and the 
energy difference $\Delta E=E_B-E_{\rm beam}$, where $E_{\rm beam}$ is the beam energy, 
and $E_B$ and $\vec{p}_B$ are the reconstructed energy and momentum, respectively, of 
the $\Bz$ candidate. All quantities are evaluated in the CM frame. 
Events satisfying $M_{\rm bc}>5.24 \gevcc$ and $-0.20 \gev <\Delta E <0.10 \gev$ 
are retained for further analysis. To calculate the signal yield, we define a 
smaller signal region: $5.27~\gevcc<M_{\rm bc} < 5.29 \gevcc$ and 
$-0.10~\gev <\Delta E< 0.05 \gev$. In order to suppress ``continuum'' 
background arising from light quark production ($e^+e^-\to\qqbar,\,q=u,d,s,c)$, 
we require that the event shape variable $R_2$, which is the ratio 
of second to zeroth Fox-Wolfram moments~\cite{FW}, satisfies $R_2<0.4$.

After applying all selection criteria, 2.9\% of events have multiple $\Bz$ candidates 
in the signal region. For these events, we retain the candidate having the smallest 
sum of $\chi^2$ values obtained from the $\pi^0\to\gamma\gamma$ mass-constrained fit
and the $\jpsi\to \ell^+\ell^-$  vertex- and mass-constrained fit.
According to MC simulations, this criterion selects the correct $\Bz$ candidate 
in 74\% of multiple-candidate events.


We tag (identify) the flavor of the accompanying $B$ meson using inclusive properties 
of particles not associated with the signal $\Bz\to\jpsi\piz$ decay. The 
algorithm for flavor tagging is described in Ref.~\cite{Kakuno:2004cf}. Two 
parameters, $q$ and $r$, are used to represent the tagging information. The 
former is the implied flavor of the signal $B$ decays as used in Eq.~(\ref{eq:one}). 
The latter is an event-by-event MC-determined quality factor that ranges from $r=0$ 
for no flavor discrimination to $r= 1$ for unambiguous flavor assignment. It is used 
for sorting candidate events into seven $r$ ranges. For events having $r>0.10$, 
we determine the wrong-tag fractions $\omega^{}_l$ ($l=1,7$) and their differences 
$\Delta\omega^{}_l$ between $\Bz$ and $\Bzb$ decays from 
a control sample of self-tagged semileptonic and hadronic $b\to c$ decays~\cite{Abe:2004mz, Adachi:2012et}.
If $r < 0.10$, the wrong tag fraction is set to 0.5.

The vertex position for the $\Bz\to\jpsi\piz$ decay is reconstructed using lepton tracks from $\jpsi$ decays.
We perform a vertex fit with a constraint to the interaction point (IP) profile. 
A vertex position for $f_{\rm tag}$ is obtained using tracks that are not 
assigned to the  $B^0\to\jpsi\piz$ candidate, plus the IP constraint.
This constraint allows for reconstruction of an $f_{\rm tag}$ vertex even 
in cases when only one track candidate satisfies the requirement on SVD hits.
The fraction of single-track vertices for $f_{\rm tag}$ is approximately 12\%, estimated from MC.
To reject events with poorly reconstructed vertices, we require 
$\sigma_z<200~\mu$m and $h<50$ for multi-track vertices, and 
$\sigma_z<500~\mu$m for single-track vertices, where $\sigma_z$ is 
the error on the vertex $z$ coordinate, and $h$ is the $\chi^2$ value
calculated in three-dimensional space without using the IP 
constraint~\cite{Adachi:2012et}. We retain events in which 
both $\jpsi$ and $f^{}_{\rm tag}$ vertices are reconstructed 
and satisfy $|\Delta t|<70$~ps.


To extract the signal yield, we perform a two dimensional 
unbinned maximum likelihood fit to the variables 
$M_{\rm bc}$ and $\Delta E$.
The probability density function (PDF) of signal events consists of two parts: one for candidates that are 
correctly reconstructed, and one for those incorrectly reconstructed, 
$i.e.$, at least one  daughter originates from the other (tag-side) $B$. 
For the former case, both the $M_{\rm bc}$ 
and $\Delta E$ distributions are modeled with Crystal Ball (CB) 
functions~\cite{Skwarnicki:1986xj}. For the latter case, the 
correlated two-dimensional $M_{\rm bc}$-$\Delta E$ distribution 
is modeled with a non-parametric PDF~\cite{Cranmer:2000du}.
The fraction of incorrectly reconstructed  decays ($\sim10\%$ in the signal region) is taken from MC simulation. 
The CB parameters that describe the lower tail of the $M_{\rm bc}$ and $\Delta E$ 
distributions are also fixed to MC values. 

\begin{figure}[h!t!p!]
\begin{center}
    \includegraphics[width=0.45\textwidth]{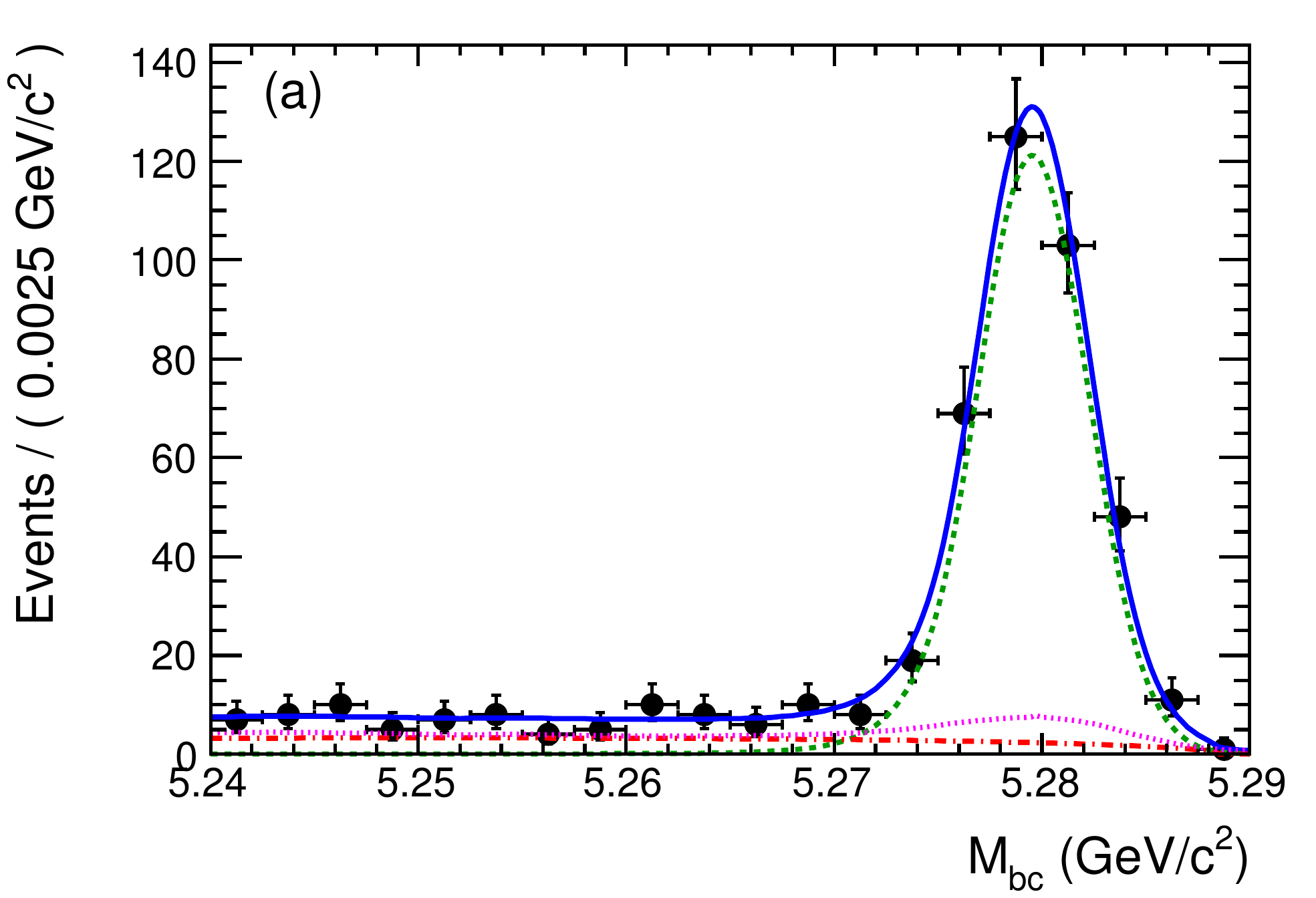}
    \includegraphics[width=0.45\textwidth]{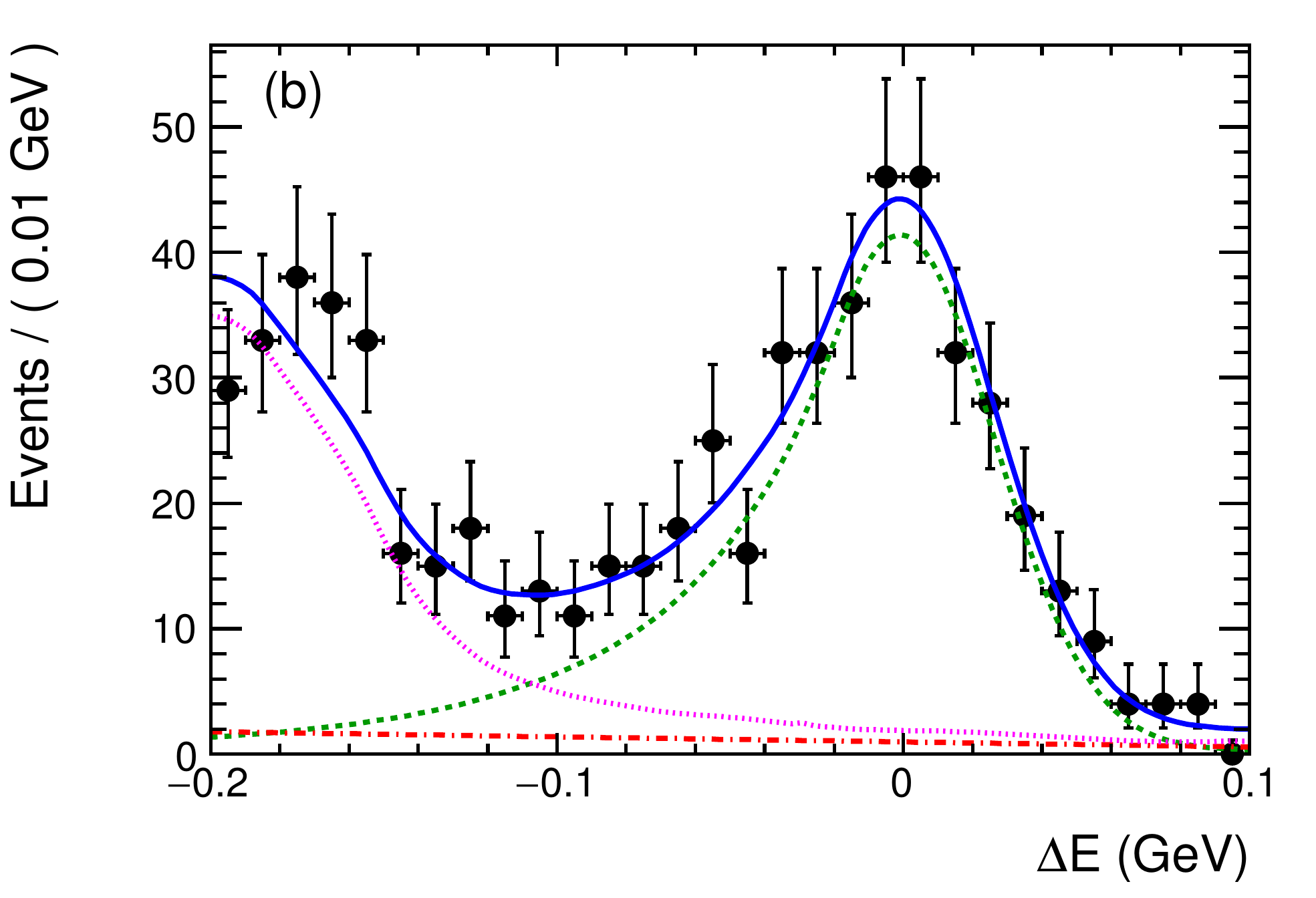}
\end{center}\label{fig:svda_mbc_de_full}
\vskip -0.65cm
\caption{\small Projections of the two-dimensional fit:  
(a) $M_{\rm bc}$ in the $\Delta E$ signal region, and 
(b) $\Delta E$ in the $M_{\rm bc}$ signal region. The points are data, 
the (green) dashed curves show the signal, the (red) dot-dashed curves show the 
$q\bar{q}$ background, the (magenta) dotted curves show the $\BBbar$ background, 
and the (blue) solid curves show the total PDF.}
\end{figure}

The remaining background is small and dominated by $\BBbar$ events in which 
one of the $B$ mesons decays into a final state containing a $\jpsi$. We divide
this background into three categories: 
(a) $\Bz\to\jpsi\KS$, (b) $\Bz\to\jpsi\KL$, and (c) $B\to\jpsi X$ other than 
$\Bz\to\jpsi\Kz$. 
We use two-dimensional non-parametric PDFs~\cite{Cranmer:2000du} to model the 
$M_{\rm bc}$-$\Delta E$ distributions for all three categories. We fix the  
background yields to those expected based on MC simulation:
10.8 $\jpsi\KS$ events, 10.0 $\jpsi\KL$ events, and 17.5 other $\jpsi X$ 
events in the $M_{\rm bc}$-$\Delta E$ signal region. The remaining 
background  comes from continuum $\qqbar$ events. 
We model the $M_{\rm bc}$ and $\Delta E$ distributions of continuum background 
with an ARGUS~\cite{Albrecht:1990am} function having its endpoint fixed to $5.29~\gevcc$, and a 
first-order polynomial, respectively.  Background coming from $\BBbar$ not containing 
a real $\jpsi$ is negligible. 
From the fit we obtain $330.2\pm22.1$ signal events  and $16.3\pm 3.5$ 
continuum events. The purity of the signal is 86\% in the signal region. 
Projections of the fit are shown in Fig.~\ref{fig:svda_mbc_de_full}. 

The branching fraction is calculated from the formula
\begin{eqnarray}
 \BF(\Bz\to\jpsi\piz) =  \frac{Y_{\rm sig}}{\varepsilon\times N_{\BBbar}\times \mathcal{B}_{\jpsi}\times\mathcal{B}_{\piz}}\,,
\label{eq:br}
\end{eqnarray}
where $Y_{\rm sig}$ is the fitted signal yield; 
$N_{\BBbar}=(772\pm11)\times10^6$ is the number of $\BBbar$ events;  
$\varepsilon=(22.3\pm0.1)\%$ is the signal efficiency (for $\epem$ and $\mup\mun$ combined) as obtained from MC simulation; 
$\BF_{\jpsi}$ is the sum of $\BF(\jpsi\to\mup\mun)$ and
$\BF(\jpsi\to\epem)$~\cite{PDG}; and $\BF_{\piz}$ is the branching fraction of $\piz\to\gamma\gamma$~\cite{PDG}.
In Eq.~(\ref{eq:br}) we assume equal production of $\Bz\Bzb$ and 
$\Bp\Bm$ pairs at the \FourS resonance.  The result is
\begin{eqnarray*}
\BF(\Bz\to\jpsi\piz)=(1.62\pm0.11\pm0.06)\times10^{-5},
\end{eqnarray*}
where the first uncertainty is statistical and the second is systematic.

The systematic uncertainty on $\BF(\Bz\to\jpsi\piz)$ arises from several sources, 
as listed in Table~\ref{sys1}. The uncertainty due to the fixed parameters in the PDF 
is estimated by varying each parameter individually according to its statistical 
uncertainty. 
The resulting changes in the branching fraction are added in quadrature and the result 
is taken as the systematic uncertainty. 
The non-parametric shapes are also varied by changing their smoothing, and the associated 
systematic uncertainty is found to be negligible.  
We assign a 1.5\% systematic uncertainty due 
to $\piz$ reconstruction, as determined from a study of $\tau^-\to\pi^-\piz\nu_{\tau}$ 
decays~\cite{Ryu:2014vpc}. The uncertainty due to charged track reconstruction is 0.35\% per track, 
as determined from a study of partially reconstructed $D^{*+}\to D^0\pip,\Dz\to\KS\pip\pim$ 
decays. We assign a 2.1\% uncertainty due to lepton identification, as obtained  from a 
study of two-photon~$\gamma\gamma\to\ell^+\ell^-$ production events.
The uncertainty due to the estimated fraction of incorrectly reconstructed signal 
events is obtained by varying this fraction by $\pm 100\%$. 
As $\B\to\jpsi(\KS,\KL,X)$ decays are well measured, we evaluate the uncertainty due to their estimated amounts by varying them by $\pm 20\%$.
The uncertainty due to the number of $\BBbar$ pairs is 1.4\%, and the 
uncertainty on the reconstruction efficiency $\varepsilon$ due to the MC sample 
size is 0.4\%.  
The total systematic uncertainty  is obtained by summing all individual 
contributions in quadrature.

\begin{table}[htb]
\renewcommand{\arraystretch}{1.2}
\caption{\small Fractional systematic uncertainties for $\mathcal{B}(\Bz\to\jpsi\pi^0)$.}
\label{sys1}
\centering
\begin{tabular}{l|c}
\hline \hline
Source & Uncertainty (\%) \\
\hline
PDF parametrization & $0.1$\\
$\pi^0$ reconstruction & 1.5 \\
Tracking  & 0.7 \\
Lepton-ID selection & 2.1\\
Incorrectly reconstructed signal events & 0.8 \\
$\B\to\jpsi(\KS,\KL,X)$ background & $^{+1.8}_{-2.0}$\\
MC statistics & 0.4 \\
Secondary branching fractions  & 0.8 \\
Number of $B\Bbar$ pairs & 1.4\\
\hline
Total & $^{+3.7}_{-3.9}$\\
\hline\hline
\end{tabular}
\end{table}


We determine $\mathcal{S}$ and $\mathcal{A}$ by performing an unbinned maximum 
likelihood fit to the $\Delta t$ distribution  of candidate events in the signal 
region. The PDF for the signal component, 
$\mathcal{P}_{\rm sig}(\Delta t; \mathcal{S}, \mathcal{A}, q, \omega_l, \Delta \omega_l)$, 
is given by Eq.~(\ref{eq:one}) with the parameters $\tau_{\Bz}$ and $\Delta m_d$ 
fixed to the world-average values~\cite{Amhis:2014hma}. We modify this expression 
to take into account the effect of incorrect flavor assignment, which is 
parametrized by $\omega_l$ and $\Delta \omega_l$. This PDF is then convolved 
with the decay-time resolution function $R_{\rm sig}(\Delta t)$. The resolution 
function is itself a convolution of four components: 
the detector resolutions for $z^{}_{\jpsi\piz}$ and $z^{}_{\rm tag}$; 
the shift of the $z_{\rm tag}$ vertex position due to secondary tracks from 
charmed particle decays; and the kinematic approximation that the $\B$ mesons 
are at rest in the CM frame~\cite{Adachi:2012et}. The PDFs for $\Bz\to\jpsi\KS$ 
and $\Bz\to\jpsi\KL$ backgrounds are the same as $\mathcal{P}_{\rm sig}$
but with \CP parameters $\mathcal{A}$ and $\mathcal{S}$ fixed to the recent 
\belle results~\cite{Adachi:2012et}. The PDF for $\B\to\jpsi X$ background 
is taken to have the same form as $\mathcal{P}_{\rm sig}$ but with $\mathcal{A}$ 
and $\mathcal{S}$ set to zero, and with an effective lifetime $\tau^{}_{\rm eff}$ 
determined from MC simulation. The PDF for continuum background is taken to
be the sum of two Gaussian functions whose parameters are obtained by fitting events 
in the sideband region $5.20 \gevcc < M_{\rm bc} < 5.26 \gevcc$ and 
$0.10 \gev < \Delta E < 0.50 \gev$.  

\begin{figure}[h!t!p!]
\begin{center}
    \includegraphics[width=0.45\textwidth]{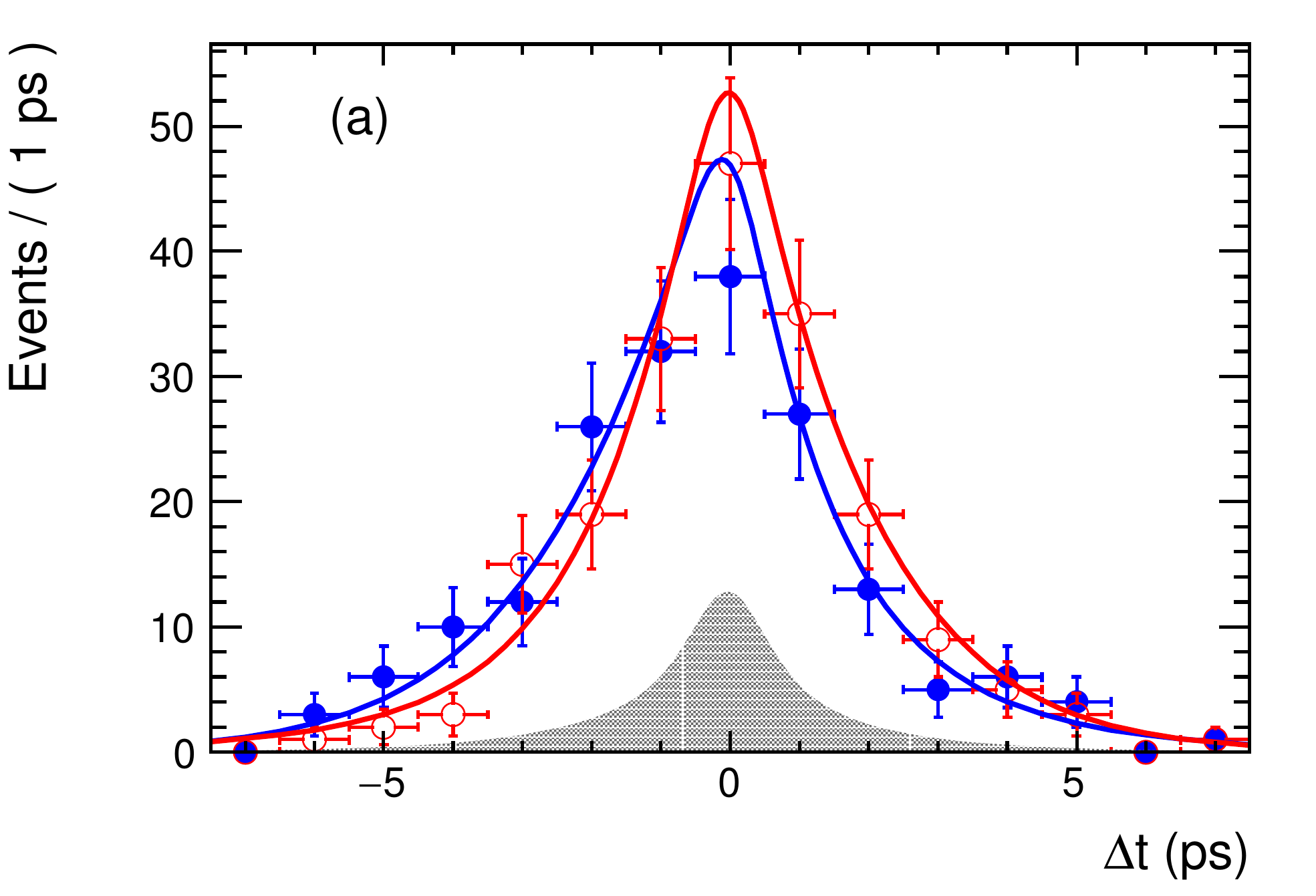}
        \includegraphics[width=0.45\textwidth]{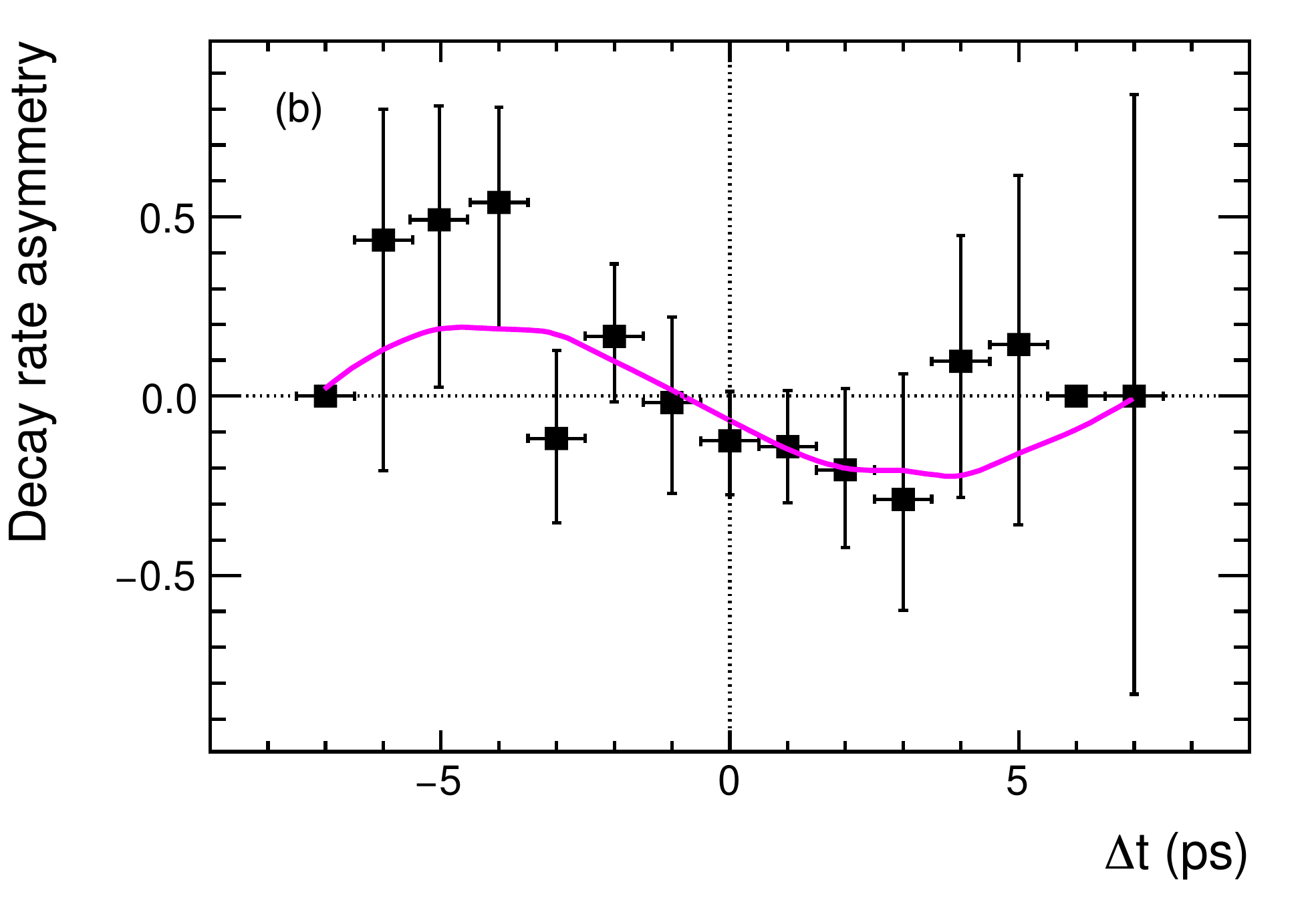}%
\end{center}\label{fig:svda_dt_full}
\vskip -0.65cm
\caption{\small (a) Distributions of  $\Delta t$. The (blue) solid and (red) open 
points represent the $q=+1$ and $q=-1$  events, respectively, and the solid curves 
show the corresponding fit projections. The gray shaded region represents the sum of all
backgrounds.  (b) Time-dependent asymmetry ${\cal A}^{}_{CP}$ (see text). }
\end{figure}

We assign the following likelihood for the $i$-th event:
\begin{linenomath}
\begin{eqnarray}
\mathcal{P}_i(\Delta t) & = & 
(1-f_{\rm ol})\int d(\Delta t') \Big[ R_{\rm sig}(\Delta t_i-\Delta t')\times  \nonumber \\
 & & \Big( f_{\rm sig}\mathcal{P}_{\rm sig}(\Delta t') + f_{\jpsi\KS}\mathcal{P}_{\jpsi\KS}(\Delta t') \nonumber \\
 & + & f_{\jpsi\KL}\mathcal{P}_{\jpsi\KL}(\Delta t') + f_{\jpsi X}\mathcal{P}_{\jpsi X}(\Delta t') 
\Big)  \nonumber \\
 & + &  f_{q\bar{q}}\mathcal{P}_{q\bar{q}}(\Delta t_i) \Big] +  f_{\rm ol}\mathcal{P}_{\rm ol}(\Delta t_i)\,,
\end{eqnarray} 
\end{linenomath}
where $f_{\rm sig}$, $f_{\jpsi\KS}$, $f_{\jpsi\KL}$, $f_{\jpsi X}$, and $f_{q\bar{q}}$ are 
the fractions of signal, $\Bz\to\jpsi\KS$, $\Bz\to\jpsi\KL$, $\B\to\jpsi X$, and $q\bar{q}$
continuum background, respectively. All fractions depend on the flavor tagging quality $r$ 
and are functions of $\Delta E$ and $M_{\rm bc}$. The term $\mathcal{P}_{\rm ol}(\Delta t)$ 
is a broad Gaussian function that represents an outlier component with a small fraction 
$f_{\rm ol}\approx0.5\%$. The only free parameters in the  fit are $\mathcal{S}$ and $\mathcal{A}$; 
these are determined by maximizing the likelihood 
$\mathcal{L}(\mathcal{S}, \mathcal{A}) = 
\prod_i\mathcal{P}_i(\Delta t_i; \mathcal{S}, \mathcal{A})$. 
Figure~\ref{fig:svda_dt_full} shows the fitted $\Delta t$ distribution and the 
time-dependent decay rate asymmetry $\mathcal{A}_{\CP}$, where 
$\mathcal{A}_{\CP}=
\bigl( Y^{(q=+1)}_{\rm sig} - Y^{(q=-1)}_{\rm sig}\bigr)/
\bigl( Y^{(q=+1)}_{\rm sig} + Y^{(q=-1)}_{\rm sig}\bigr)$, where $Y^{(q=\pm1)}_{\rm sig}$ is the signal yield with $q=\pm1$.
The results of the fit are
\begin{eqnarray*}
\mathcal{S} & = & -0.59 \pm 0.19 \pm 0.03 \\
\mathcal{A} & = & -0.15 \pm 0.14\,^{+0.04}_{-0.03}\,,
\end{eqnarray*}
where the first uncertainty is statistical and the second is systematic. 
The correlation between $\mathcal{A}$ and $\mathcal{S}$ is $-0.005$.

The systematic uncertainties for $\mathcal{S}$ and $\mathcal{A}$ are 
listed in Table~\ref{sys2}. They are small compared to the corresponding 
statistical uncertainties. 
\begin{table}[htb]
\renewcommand{\arraystretch}{1.2}
\caption{\small Absolute systematic uncertainties for $\mathcal{S}$ and $\mathcal{A}$.}
\label{sys2}
\centering
\begin{tabular}{l|cc}
\hline \hline
Source &  $\sigma_{\mathcal{S}}$~(\%) & $\sigma_{\mathcal{A}}$~(\%)  \\
\hline
Vertex reconstruction & $^{+2.36}_{-1.75}$ & $^{+1.40}_{-2.22}$\\
Resolution function &  $^{+1.43}_{-2.37}$ & $^{+1.00}_{-0.91}$\\
Physics parameters & $^{+0.04}_{-0.03}$ & $\pm 0.04$\\
Fit bias & $\pm0.68$ & $\pm 0.27$\\
Wrong tag fraction & $^{+0.41}_{-0.20}$ & $^{+0.43}_{-0.17}$ \\
$M_{\rm bc}$, $\Delta E$ shapes & $^{+0.52}_{-0.45}$ & $^{+0.50}_{-0.48}$\\
Signal and background fraction & $^{+0.71}_{-0.62}$ & $^{+0.49}_{-0.72}$\\
Background $\Delta t$ shape & $^{+0.20}_{-0.12}$ & $\pm0.10$\\
Tag-side interference & $\pm0.20$ & $^{+3.80}_{-0.00}$ \\
\hline
Total & $^{+3.02}_{-3.14}$ &$^{+4.26}_{-2.57}$\\
\hline\hline
\end{tabular}
\end{table}
The largest contributions to $\mathcal{S}$ arise from vertex reconstruction 
and the resolution function. The uncertainty due to the former includes 
uncertainties in the IP profile, charged track selection, vertex quality selection,
and SVD misalignment. We vary each parameter of the resolution function 
by one standard deviation ($\pm 1\sigma$) and compare the resulting fit result with 
that of the nominal fit; the difference between the two is taken as the systematic uncertainty.
Each physics parameter  that is fixed to its
world average value~\cite{Amhis:2014hma}, $e.g.$, $\tau_{B^0}$ and $\Delta m_d$,  is varied by the corresponding error;
the uncertainty is taken to be the resulting difference with the nominal 
fit result. 
The uncertainty due to possible fit bias is evaluated using large ensembles 
of MC signal events; the differences of the fit results with the MC inputs 
are assigned as systematic uncertainties.
The uncertainties due to $\omega^{}_l$ and $\Delta\omega^{}_l$ are estimated by varying 
these parameters individually by $\pm 1\sigma$.  The $M_{\rm bc}$ and $\Delta E$ shape 
parameters, and the fractions of signal and background, are varied 
to  estimate their contributions to the systematic uncertainty. We vary each parameter in 
$\mathcal{P}_{q\bar{q}} (\Delta t)$ and $\mathcal{P}_{\jpsi X} (\Delta t)$ by $\pm 1\sigma$.  
For $\mathcal{P}_{\jpsi \KS} (\Delta t)$ and $\mathcal{P}_{\jpsi \KL} (\Delta t)$, we vary 
the \CP asymmetry parameters by their statistical errors~\cite{Adachi:2012et}.
We include the effect of tag-side interference~\cite{Long:2003wq}, which introduces 
a significant contribution to the systematic uncertainty for $\mathcal{A}$. 
Tag-side interference is caused by interference between the two tree-level 
amplitudes contributing to $B\to DX$ decays.


In summary, we have measured the branching fraction and time-dependent \CP asymmetry 
for $\Bz\to\jpsi\piz$ decays using the full \belle \FourS data set. The results are
\begin{eqnarray*}
\mathcal{B} & = & (1.62 \pm 0.11 \pm0.06)\times10^{-5} \\
\mathcal{S} & = & -0.59 \pm 0.19 \pm 0.03  \\
\mathcal{A} & = & -0.15 \pm 0.14\,^{+0.04}_{-0.03}\,,
\end{eqnarray*}
where the first uncertainty is statistical and the second is systematic. 
The measured value for the branching fraction is the most precise value to date and supersedes
the previous measurement~\cite{Abe:2002rc}. It 
is consistent with measurements made by other experiments~\cite{Aubert:2008bs,  Avery:2000yh}. The measured \CP asymmetries 
are consistent with, and supersede, our previous results~\cite{Lee:2007wd}. 
The direct \CP asymmetry $\mathcal{A}$ is consistent with zero.
The mixing-induced \CP asymmetry $\mathcal{S}$ differs from zero ($i.e.$, no \CP violation) 
by 3.0$ \sigma$, and it differs from the BaBar result~\cite{Aubert:2008bs}
(which is outside the physical region) 
by 3.2$\sigma$. The value  is consistent 
with the value of $\sin 2\phi_1$ measured using $b\to\ccbar s$ decays~\cite{PDG}. These results
indicate that the penguin and any NP contribution to $\Bz\to\jpsi\piz$ 
are small.

\begin{center}
\textbf{ACKNOWLEDGMENTS}
\end{center}
We thank the KEKB group for the excellent operation of the
accelerator; the KEK cryogenics group for the efficient
operation of the solenoid; and the KEK computer group, and the Pacific Northwest National
Laboratory (PNNL) Environmental Molecular Sciences Laboratory (EMSL)
computing group for strong computing support; and the National
Institute of Informatics, and Science Information NETwork 5 (SINET5) for
valuable network support.  We acknowledge support from
the Ministry of Education, Culture, Sports, Science, and
Technology (MEXT) of Japan, the Japan Society for the 
Promotion of Science (JSPS), and the Tau-Lepton Physics 
Research Center of Nagoya University; 
the Australian Research Council including grants
DP180102629, 
DP170102389, 
DP170102204, 
DP150103061, 
FT130100303; 
Austrian Science Fund under Grant No.~P 26794-N20;
the National Natural Science Foundation of China under Contracts
No.~11435013,  
No.~11475187,  
No.~11521505,  
No.~11575017,  
No.~11675166,  
No.~11705209;  
Key Research Program of Frontier Sciences, Chinese Academy of Sciences (CAS), Grant No.~QYZDJ-SSW-SLH011; 
the  CAS Center for Excellence in Particle Physics (CCEPP); 
the Shanghai Pujiang Program under Grant No.~18PJ1401000;  
the Ministry of Education, Youth and Sports of the Czech
Republic under Contract No.~LTT17020;
the Carl Zeiss Foundation, the Deutsche Forschungsgemeinschaft, the
Excellence Cluster Universe, and the VolkswagenStiftung;
the Department of Science and Technology of India; 
the Istituto Nazionale di Fisica Nucleare of Italy; 
National Research Foundation (NRF) of Korea Grants
No.~2015H1A2A1033649, No.~2016R1D1A1B01010135, No.~2016K1A3A7A09005
603, No.~2016R1D1A1B02012900, No.~2018R1A2B3003 643,
No.~2018R1A6A1A06024970, No.~2018R1D1 A1B07047294; Radiation Science Research Institute, Foreign Large-size Research Facility Application Supporting project, the Global Science Experimental Data Hub Center of the Korea Institute of Science and Technology Information and KREONET/GLORIAD;
the Polish Ministry of Science and Higher Education and 
the National Science Center;
the Grant of the Russian Federation Government, Agreement No.~14.W03.31.0026; 
the Slovenian Research Agency;
Ikerbasque, Basque Foundation for Science, Basque Government (No.~IT956-16) and
Ministry of Economy and Competitiveness (MINECO) (Juan de la Cierva), Spain;
the Swiss National Science Foundation; 
the Ministry of Education and the Ministry of Science and Technology of Taiwan;
and the United States Department of Energy and the National Science Foundation.

\end{document}